\documentclass[reprint,amsmath,amssymb,aps,prb]{revtex4-2}
\usepackage{graphicx}
\usepackage{dcolumn}
\usepackage{bm}
\usepackage{comment}
\usepackage{amsmath, amssymb}
\usepackage{bbold}
\usepackage[usenames,dvipsnames,svgnames,table]{xcolor}
\usepackage{makecell}
\usepackage[colorinlistoftodos,prependcaption,textsize=tiny]{todonotes}
\usepackage{float}
\usepackage{tabularx}
\usepackage{hyperref}

\newcommand{\g}[2]{\tau_{#1} \otimes \sigma_{#2}}
\newcommand{\gp}[2]{$\tau_{#1} \otimes \sigma_{#2}$}

\newcommand{\ignore}[1]{}

\hypersetup{
    colorlinks,
    citecolor=black,
    filecolor=black,
    linkcolor=blue,
    urlcolor=blue
}

\allowdisplaybreaks

\begin{document}

\preprint{APS/123-QED}

\title{Composite superconducting orders and magnetism in CeRh\textsubscript{2}As\textsubscript{2}}

\author{Fabian Jakubczyk}
\email{fabian.jakubczyk@tu-dresden.de}
\affiliation{Institute of Theoretical Physics and Würzburg--Dresden Cluster of Excellence ctd.qmat, Technische Universität Dresden, 01069 Dresden}

\author{Julia M. Link}
\affiliation{Institute of Theoretical Physics and Würzburg--Dresden Cluster of Excellence ctd.qmat, Technische Universität Dresden, 01069 Dresden}

\author{Carsten Timm}
\email{carsten.timm@tu-dresden.de}
\affiliation{Institute of Theoretical Physics and Würzburg--Dresden Cluster of Excellence ctd.qmat, Technische Universität Dresden, 01069 Dresden}

\date{\today}

\begin{abstract}

Locally noncentrosymmetric materials are attracting significant attention due to the unique phenomena associated with sublattice degrees of freedom. The recently discovered heavy-fermion superconductor CeRh\textsubscript{2}As\textsubscript{2} has emerged as a compelling example of this class, garnering widespread interest for its remarkable temperature-magnetic-field phase diagram, which features a field-induced first-order superconductor-to-superconductor phase transition with nontrivial dependence on the field direction and high critical fields, as well as antiferromagnetic and potentially higher multipole orders. To investigate the complex interplay of the ordered phases in CeRh\textsubscript{2}As\textsubscript{2}, we develop a theoretical framework based on symmetry analysis applied to a Bogoliubov--de Gennes Hamiltonian and Landau methods. This approach allows us to propose probable symmetries of the superconducting states and elucidate their close relationship with magnetism. Among other results, we find that the near degeneracy of two pairing symmetries is naturally explained if and only if intralayer spin-orbit coupling is large compared to interlayer hopping. Intriguingly, we find that the first-order transition can be interpreted as a transition between coexistence phases of the same superconducting order parameters, albeit with distinct admixtures. This line may end in a critical end point below the superconducting critical temperature. Our approach accurately reproduces current experimental phase diagrams for varying temperature as well as out-of-plane and in-plane magnetic field, both if the transition to a magnetic phase occurs below the superconducting critical temperature and if it occurs above. Furthermore, we calculate the magnetic susceptibility and the specific heat and compare these quantities to recent experimental results.

\end{abstract}

\maketitle

\section{Introduction}
\label{sec:intro}

Layered materials that overall possess inversion symmetry while the separate layers do not have this symmetry show intriguing superconducting phenomena enabled by the sublattice degrees of freedom \cite{FLS11, MSY12, SAF14, YSY14, fischer_superconductivity_2023}. A particularly interesting example is the heavy-fermion material CeRh\textsubscript{2}As\textsubscript{2}, which has recently received attention due to its remarkable multiphase superconductivity \cite{khim_field-induced_2021}. The material becomes superconducting with reported critical temperatures ranging from $T_c \approx 0.26\,\mathrm{K}$ \cite{khim_field-induced_2021} to $0.37\,\mathrm{K}$ \cite{kibune_observation_2022}, in both cases obtained from the susceptibility. A magnetic field applied along the crystallographic \textit{z}-axis suppresses $T_c$ but the suppression is weaker than expected for Pauli limiting in a conventional superconductor \cite{khim_field-induced_2021}. A key feature is the field-induced first-order transition between different superconducting orders at about $4\,\mathrm{T}$, which depends only weakly on temperature \cite{khim_field-induced_2021}. The high-field phase is only very weakly suppressed when the field is further increased and thus evidently avoids Pauli limiting \cite{khim_field-induced_2021}, with an extrapolated upper critical field at zero temperature of about $14\,\mathrm{T}$. The first-order transition is widely interpreted as an even-to-odd-parity transition, with a potential (pseudo-) spin-triplet high-field superconducting state \cite{khim_field-induced_2021, landaeta_field-angle_2022, cavanagh_nonsymmorphic_2022, ogata_parity_2023, NaB24}.

On the other hand, a magnetic field applied in the \textit{ab} plane suppresses the low-field superconducting phase at an upper critical field of $1.9\,\mathrm{T}$, consistent with Pauli limiting \cite{khim_field-induced_2021}. No transition to another superconducting phase is observed for a magnetic field in plane. However, additional phase transitions are clearly seen in several observables at stronger field and higher temperatures \cite{hafner_possible_2022}. A ``phase I'' completely surrounds the superconducting phase in the temperature-magnetic-field plane. Its critical temperature $T_0$ grows with increasing field strength within the plane \cite{hafner_possible_2022, khanenko_origin_2025}. The phase gives way to a ``phase II'' at a first-order transition at a weakly temperature-dependent field of about $9\,\mathrm{T}$ and the critical temperature $T_0$ further increases with field \cite{hafner_possible_2022, khanenko_origin_2025}. The phases I and II were suggested to exhibit quadrupolar order since they are not visible in the magnetic dipolar response and because the enhancement of $T_0$ with magnetic field is hard to understand for dipolar order~\cite{hafner_possible_2022, mishra_anisotropic_2022}.

Beyond exhibiting multiple superconducting phases, CeRh\textsubscript{2}As\textsubscript{2} also demonstrates the coexistence of magnetism and superconductivity. On the one hand, $^{75}\mathrm{As}$ nuclear quadrupole resonance (NQR) \cite{kibune_observation_2022} and nuclear magnetic resonance (NMR) \cite{ogata_parity_2023, ogata_appearance_2024} exhibit an onset of magnetic order below the superconducting transition. On the other hand, recent muon spin relaxation ($\mu$SR) data \cite{khim_coexistence_2024} suggest that phase I, which sets in at $T_0 \approx 0.55\,\mathrm{K}$, above the superconducting critical temperature of $T_c \approx 0.3\,\mathrm{K}$, in fact involves magnetic dipolar order.

In this work, we study the interplay between competing orders in CeRh\textsubscript{2}As\textsubscript{2} based on symmetry analysis, plausible physical assumptions, and free-energy expansion in terms of multiple order parameters (OPs). Our approach shows that the phase diagram down to rather detailed aspects does not depend on details of the electronic structure and the shape of the Fermi surface, which are not well known. Recent angle-resolved photoelectron spectroscopy results \cite{WZJ24, chen_coexistence_2024} still have limited resolution and are performed at significantly higher temperatures than relevant here. Results of ab-initio calculations tend to depend rather strongly on details of the method and on the interaction strength \cite{Zwi92, Zwi16, NDI21, ptok_electronic_2021, cavanagh_nonsymmorphic_2022, ishizuka_correlation-induced_2024}. On the other hand, this approach allows us to identify those aspects of the electronic structure that do have a large impact on the phase diagram. We identify the symmetry of the dominant superconducting pairing states as well as the magnetic phase. We find that the low-field phase is dominated by even-parity, spin-singlet $B_{1g}$ pairing, whereas the high-field phase shows predominant odd-parity $B_{1u}$ and $B_{2u}$ contributions. The magnetic OP likely has time-reversal-odd (TR-odd) $A_{1u}$ symmetry and is antiferromagnetic (AFM). However, the situation is more complex since we find an admixture of four superconducting OPs of different symmetries and the AFM OP in various regions of realistic phase diagrams. This is accompanied by the novel observation of a symmetry-preserving first-order transition between coexistence phases of the same overall symmetry, yet with distinct proportions of the individual superconducting OPs. Moreover, the symmetry-preserving first-order transition allows and we indeed predict that the first-order line ends in a critical end point. Beyond this point, the low-field and high-field phases are connected by a crossover region. Our approach coherently reproduces state-of-the-art experimental results for both orderings of the superconducting critical temperature $T_c$ and the N\'eel temperature of the magnetic order, $T_N > T_c$ and $T_N < T_c$, while introducing the important perspective of not just multiphase, but also multicomponent superconductivity in CeRh\textsubscript{2}As\textsubscript{2}. These multicomponent phases involve nondegenerate composite OPs of distinct irreducible representations and amplitudes, as opposed to the case of OPs that belong to the same irreducible representation and are degenerate by symmetry.

The remainder of this paper is structured as follows: In Sec.\ \ref{sec:model}, we introduce our symmetry-informed model for the normal state of CeRh\textsubscript{2}As\textsubscript{2}. In Sec.\ \ref{sec:SC_pairing_states}, we then analyze which superconducting pairing states might exist and which are most likely in the absence of an applied magnetic field. Section \ref{sec:magnetism} discusses the magnetic order observed in CeRh\textsubscript{2}As\textsubscript{2} and its coexistence with superconductivity. In Sec.\ \ref{sec:field_induced}, we investigate the effect of a magnetic field on the superconducting order, depending on the field direction. Section \ref{sec:Landau_expansion} first introduces the Landau free-energy expansion for our model and then presents our main results for phase diagrams in the temperature-magnetic-field plane for various magnetic-field directions and both cases $T_N > T_c$ and $T_N < T_c$. We also show plots of the OPs and the free energy vs.\ magnetic field or temperature for cuts through the phase diagrams and discuss their interpretation. In Sec.\ \ref{sec:thermodynamics}, we briefly discuss thermodynamic signatures visible in the susceptibility and specific heat for $T_N > T_c$ and a field out of plane. We summarize our work and draw conclusions in Sec.~\ref{sec:summary}.

\section{Model and normal state}
\label{sec:model}

\begin{figure}
\centering
\includegraphics[width=.47\textwidth]{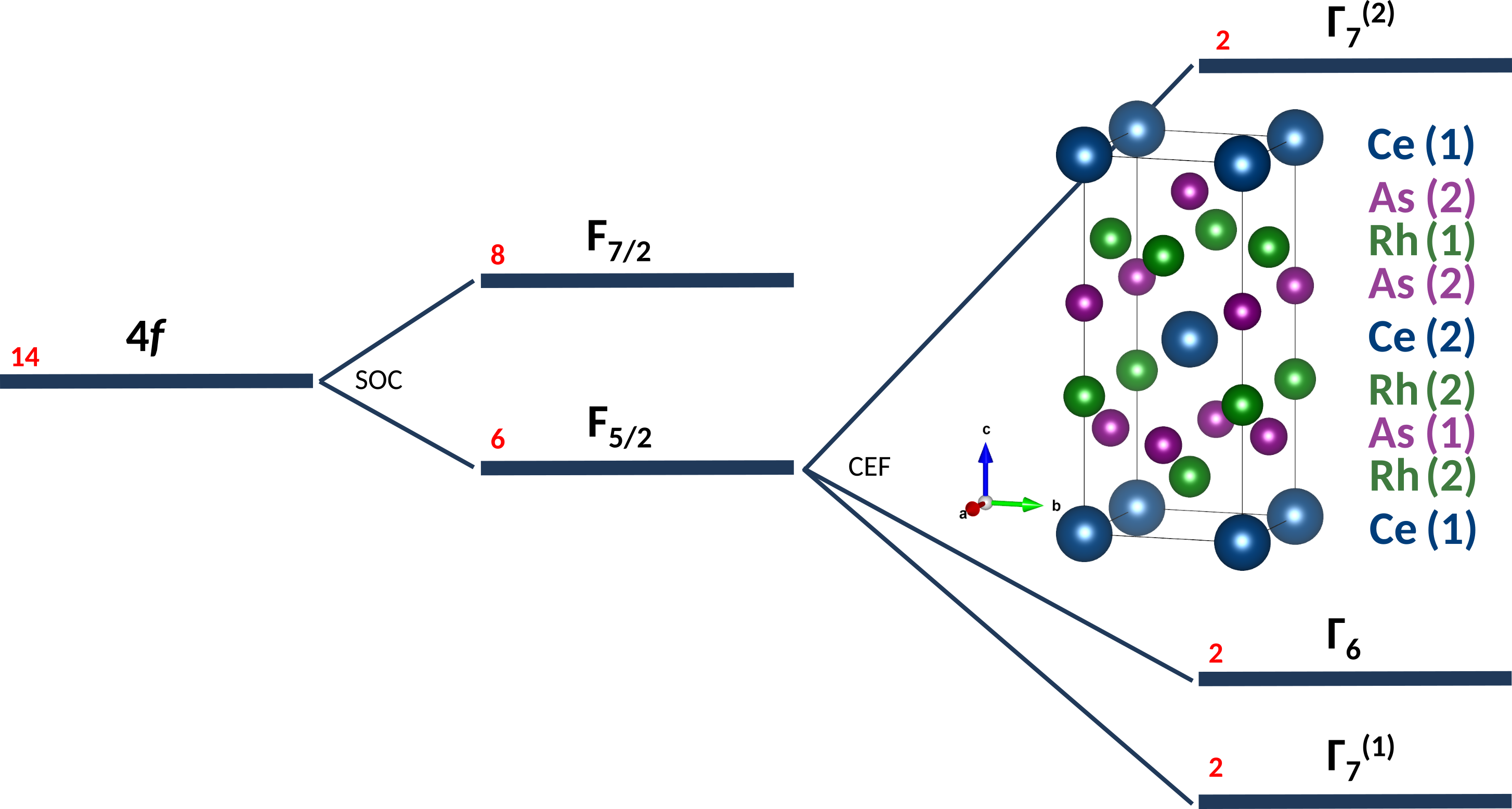}
\caption{Schematic of the CeRh\textsubscript{2}As\textsubscript{2} level splitting with the inset showing the crystal structure with distinct lattice planes. The \textit{f}-orbital degeneracy is lifted by strong spin-orbit coupling (SOC) and the $J = 5/2$ multiplet of Ce$^{3+}$ is split by the crystal electric field (CEF) into three Kramers doublets~\cite{khim_field-induced_2021, hafner_possible_2022}.}
\label{fig:level_splitting}
\end{figure}

CeRh\textsubscript{2}As\textsubscript{2} is a layered compound with $\mathrm{CaBe_2Ge_2}$-type crystal structure, nonsymmorphic space group $P4/nmm$ and point group $D_{4h}$. The structure contains square lattices of $\mathrm{Ce}$ atoms and alternating inequivalent spacer layers, as shown in Fig.~\ref{fig:level_splitting}. The unit cell contains two $\mathrm{Ce}$ atoms in adjacent layers. The inversion centers are located between the $\mathrm{Ce}$ layers, whereas there are no inversion centers within the layers.

We use an effective model with a single doublet per $\mathrm{Ce}$ atom, resulting in a $4\times 4$ normal-state Hamiltonian. The Hamiltonian can be expanded into Kronecker products \gp{i}{j}, where $\tau_x$, $\tau_y$, and $\tau_z$ are Pauli matrices and $\tau_0$ is the identity matrix acting on the basis site (sublattice) \footnote{The $\tau$ matrices act on the space spanned by the Wannier functions $\psi_1$, $\psi_2$ centered at the two Ce sites in the unit cell. Hence, $\tau_z$ leaves one unchanged and flips the sign of the other, $\tau_x$ interchanges the two, and $\tau_y$ interchanges the two with additional phase factors of $\pm i$.}, while $\sigma_x$, $\sigma_y$, and $\sigma_z$ are Pauli matrices and $\sigma_0$ is the identity matrix acting on the spin. The ground states likely form a $\Gamma_7$ ($E_{3/2}$) doublet of the local point group $C_{4v}$ at the $\mathrm{Ce}$ site \cite{hafner_possible_2022}. The first excited doublet of $\Gamma_6$ ($E_{1/2}$) symmetry is about $30\,\mathrm{K}$ above the ground-state doublet. Since $E_{3/2} = E_{1/2} \otimes B_1$, the ground-state doublet transforms differently from a spin, i.e., from a $E_{1/2}$ doublet, and we assume $E_{3/2}$ when writing down the representations of symmetry operations. However, the additional signs compared to $E_{1/2}$ drop out for bilinear forms and therefore do not affect the results. For simplicity, we refer to the $E_{3/2}$ degree of freedom as a ``spin.'' The energy splitting and the crystal structure are shown in Fig.~\ref{fig:level_splitting}.

Spatial inversion is described by
\begin{equation}
P = \g{x}{0}
\label{eq:def.P}
\end{equation}
since inversion interchanges the two Ce sites but leaves the spin invariant. Fourfold rotation about the \textit{z}-axis and twofold rotations about the \textit{x}-axis are described by
\begin{align}
C_{4z} &= \tau_0 \otimes e^{-i\pi\sigma_z/4} , \\
C_{2x} &= \tau_x \otimes e^{-i\pi\sigma_x/2} ,
\end{align}
respectively. The twofold rotation about the $[110]$ direction is then described by
\begin{equation}
C_{2xy} = C_{4z} C_{2x}
  = \tau_x \otimes e^{-i\pi(\sigma_x+\sigma_y)/2\sqrt{2}} .
\end{equation}
The twofold rotations $C_{2x}$ and $C_{2xy}$ also interchange the two Ce sites since the rotation axes lie between the layers. Finally, the antiunitary TR operator is $\mathcal{T} = U_T \mathcal{K}$ with the unitary part
\begin{equation}
U_T = \tau_0 \otimes i\sigma_y
\label{eq:def.UT}
\end{equation}
and complex conjugation $\mathcal{K}$. The matrices in sublattice and spin space are irreducible tensor operators of irreps of the gray group $D_{4h}$. The basis matrices and their irreps inferred from Eqs.\ (\ref{eq:def.P})--(\ref{eq:def.UT}) are given in Table \ref{tab:basis_matrices_model_D4h}. We are using real, i.e., orthogonal, irreps and the sign ``$\pm$'' in the subscript refers to the behavior under TR~\cite{timm_symmetry_2021}. The irreps of the $4\times 4$ basis matrices then follow from the usual rules for products of representations and are shown in Table~\ref{tab:irreps_4by4_basis_matrices}.

\begin{table}[t]
\caption{Irreps of basis matrices acting on basis-site (sublattice) space, $\tau_i$, and on spin space, $\sigma_j$. For the components of the $E_{g-}$ doublet, we take the standard convention: The first element is invariant under $C_{2x}$ and is transformed into the second by $C_{4z}$.}
\centering
\begin{tabularx}{\linewidth}{c@{\extracolsep{\fill}}c} \hline\hline
Basis matrix & Irrep \\
\hline
$\tau_0$ & $A_{1g+}$ \\
$\tau_x$ & $A_{1g+}$ \\
$\tau_y$ & $A_{2u-}$ \\
$\tau_z$ & $A_{2u+}$ \\ \hline
$\sigma_0$ & $A_{1g+}$ \\
$\begin{pmatrix} \sigma_x \\ \sigma_y \end{pmatrix}$ & $E_{g-}$ \\
$\sigma_z$ & $A_{2g-}$ \\ \hline\hline
\end{tabularx}
\label{tab:basis_matrices_model_D4h}
\end{table}

\begin{table}[b]
\caption{Irreps of $4\times4$
(\gp{i}{j}) basis matrices on combined sublattice-spin space for our model for CeRh\textsubscript{2}As\textsubscript{2}.}
\centering
\begin{tabularx}{\linewidth}{c@{\extracolsep{\fill}}c} \hline\hline
Basis matrix & Irrep \\
\hline
\gp{0}{0} & $A_{1g+}$ \\
$\begin{pmatrix} \g{0}{x} \\ \g{0}{y} \end{pmatrix}$ & $E_{g-}$ \\
\gp{0}{z} & $A_{2g-}$ \\ \hline
\gp{x}{0} & $A_{1g+}$ \\
$\begin{pmatrix} \g{x}{x} \\ \g{x}{y} \end{pmatrix}$ & $E_{g-}$ \\
\gp{x}{z} & $A_{2g-}$ \\ \hline
\gp{y}{0} & $A_{2u-}$ \\
$\begin{pmatrix} \g{y}{y} \\ -\g{y}{x} \end{pmatrix}$ & $E_{u+}$ \\
\gp{y}{z} & $A_{1u+}$ \\ \hline
\gp{z}{0} & $A_{2u+}$ \\
$\begin{pmatrix} \g{z}{y} \\ -\g{z}{x} \end{pmatrix}$ & $E_{u-}$ \\
\gp{z}{z} & $A_{1u-}$ \\ \hline \hline
\end{tabularx}
\label{tab:irreps_4by4_basis_matrices}
\end{table}

\begin{table*}[]
\caption{TR-even products of irreps of basis matrices (columns) and form factors (rows). Here, 
$X_{u+} = A_{1u+} \oplus A_{2u+} \oplus B_{1u+} \oplus B_{2u+}$ and 
$\mbox{\boldmath$X_{g+}$} = \mbox{\boldmath$A_{1g+}$} \oplus A_{2g+} \oplus B_{1g+} \oplus B_{2g+}$. Combinations invariant under the point group are highlighted by bold face.}
\centering
\begin{tabularx}{\linewidth}{l@{\extracolsep{\fill}}lllllllll} \hline \hline
Form factor & $A_{1g+}$ & $A_{2g-}$ & $E_{g-}$ & $A_{1u+}$ & $A_{1u-}$ & $A_{2u+}$ & $A_{2u-}$ & $E_{u+}$ & $E_{u-}$ \\ \hline
$A_{1g+}$ & {\boldmath$A_{1g+}$} & & & $A_{1u+}$ & & $A_{2u+}$ & & $E_{u+}$ & \\
$A_{2g+}$ & $A_{2g+}$ & & & $A_{2u+}$ & & $A_{1u+}$ & & $E_{u+}$ & \\
$B_{1g+}$ & $B_{1g+}$ & & & $B_{1u+}$ & & $B_{2u+}$ & & $E_{u+}$ & \\
$B_{2g+}$ & $B_{2g+}$ & & & $B_{2u+}$ & & $B_{1u+}$ & & $E_{u+}$ & \\
$E_{g+}$  & $E_{g+}$  & & & $E_{u+}$  & & $E_{u+}$  & & $X_{u+}$ & \\ \hline
$A_{1u-}$ & & $A_{2u+}$ & $E_{u+}$ & & {\boldmath$A_{1g+}$} & & $A_{2g+}$ & & $E_{g+}$
  \\
$A_{2u-}$ & & $A_{1u+}$ & $E_{u+}$ & & $A_{2g+}$ & & {\boldmath$A_{1g+}$} & & $E_{g+}$
  \\
$B_{1u-}$ & & $B_{2u+}$ & $E_{u+}$ & & $B_{1g+}$ & & $B_{2g+}$ & & $E_{g+}$ \\
$B_{2u-}$ & & $B_{1u+}$ & $E_{u+}$ & & $B_{2g+}$ & & $B_{1g+}$ & & $E_{g+}$ \\
$E_{u-}$  & & $E_{u+}$  & $X_{u+}$ & & $E_{g+}$ & & $E_{g+}$ & & {\boldmath$X_{g+}$}
  \\ \hline \hline
\end{tabularx}
\label{tab:form_factor_times_basis_matrices_TR_even}
\end{table*}

The normal-state Hamiltonian $H_N(\mathbf{k})$ can be expanded into these basis matrices with momentum-de\-pen\-dent form factors as
\begin{equation}
H_N(\mathbf{k}) = \sum_{n=0}^5 c_n(\mathbf{k})\, h_n ,
\label{1.HN.3}
\end{equation}
where the combination must be invariant under the point group, i.e., must belong to the trivial irrep $A_{1g+}$. Table \ref{tab:form_factor_times_basis_matrices_TR_even} shows the possible TR-even products of irreps for the basis matrices and the form factors. As noted, in $H_N(\mathbf{k})$ only combinations belonging to $A_{1g+}$ can occur, these are highlighted by bold face. The corresponding basis matrices can be read off from Table~\ref{tab:irreps_4by4_basis_matrices}~\cite{cavanagh_nonsymmorphic_2022, amin_kramers_2024}:
\begin{align}
&h_0 = \g{0}{0} && A_{1g+}, \\
&h_1 = \g{x}{0} && A_{1g+}, \\
&h_2 = \g{y}{0} && A_{2u-}, \\
&\begin{array}{@{}l}
  h_3 = \g{z}{y} \\
  h_4 = -\g{z}{x}
\end{array}\bigg\} && E_{u-}, \\
&h_5 = \g{z}{z} && A_{1u-}.
\end{align}
The form factors $c_n(\mathbf{k})$ must be basis functions of the same irreps. It is useful to collect the lowest-order basis functions compatible with the tetragonal lattice. Taking into account that the unit cell contains two $\mathrm{Ce}$ layers and that $\tau_0$ and $\tau_z$ describe processes involving the same layer (sublattice), whereas $\tau_x$ and $\tau_y$ describe processes involving different layers, the lowest-order trigonometric basis functions compatible with the nonsymmorphic space group $P4/nmm$ read as~\footnote{The matrices $h_1$ and $h_2$ contain $\tau_x$ and $\tau_y$, respectively. The corresponding form factors thus describe hopping over half-integer separations and therefore contain sine and cosine functions of half-integer multiples of $k_x$, $k_y$, $k_z$. The matrices $h_3$, $h_4$, and $h_5$ contain $\tau_z$ and the corresponding form factors contain sine and cosine functions of integer multiples of $k_x$, $k_y$, $k_z$.}
\begin{align}
&c_0(\mathbf{k}) \sim 1 && A_{1g+}, \\
&c_1(\mathbf{k}) \sim \cos\frac{k_x}{2} \cos\frac{k_y}{2}
  \cos\frac{k_z}{2} && A_{1g+}, \\
&c_2(\mathbf{k}) \sim \cos\frac{k_x}{2} \cos\frac{k_y}{2}
  \sin\frac{k_z}{2} && A_{2u-}, \\
&\begin{array}{@{}l}
  c_3(\mathbf{k}) \sim \sin k_x \\
  c_4(\mathbf{k}) \sim \sin k_y
\end{array}\bigg\} && E_{u-}, \label{1.basis.c34} \\
&c_5(\mathbf{k}) \sim \sin k_x \sin k_y \sin k_z\, (\cos k_x - \cos k_y)\! && A_{1u-}.
\label{1.basis.c5}
\end{align}
Lattice constants have been suppressed to simplify notation. The term $c_0(\mathbf{k}) h_0$ contains the chemical potential as well as hopping within the layers and between layers separated by an integer number of lattice constants (note that $\cos k_x + \cos k_y$ and $\cos k_z$ are also basis functions of $A_{1g+}$). The basis functions agree with the coefficients given in Refs.\ \cite{khim_field-induced_2021} and \cite{amin_kramers_2024}. The terms with $n=1,2$ in Eq.\ (\ref{1.HN.3}) describe hopping between neighboring $\mathrm{Ce}$ layers. The remaining terms describe spin-orbit coupling (SOC) since they contain spin Pauli matrices $\sigma_x$, $\sigma_y$, and $\sigma_z$. The leading basis functions for $n=3,4$, Eq.\ (\ref{1.basis.c34}), clearly describe in-plane nearest-neighbor coupling, which can be identified as Rashba SOC. On the other hand, the leading basis function for $n=5$, Eq.\ (\ref{1.basis.c5}), describes coupling between sites with separation vector $(1, 1, 1)$ and symmetry-related ones, where the \textit{z}-component of $1$ refers to one unit cell, i.e., two $\mathrm{Ce}$ layers. This long-range SOC is expected to be very weak~\cite{cavanagh_nonsymmorphic_2022, amin_kramers_2024}. It is of Ising type ($\sigma_z$).

The matrices $h_n$ satisfy the standard algebra of gamma matrices for inversion-symmetric models with four-valued internal degrees of freedom \cite{timm_symmetry_2021}: $h_1$, \dots, $h_5$ anticommute with each other and $h_0$ commutes with all of them. The dispersion in the normal state is thus
\begin{equation}
\xi_\pm(\mathbf{k}) = c_0(\mathbf{k}) \pm \sqrt{c_1^2(\mathbf{k}) + \ldots
  + c_5^2(\mathbf{k})} ,
\label{eq:normal_dispersion}
\end{equation}
where each band is twofold degenerate. Since $c_1(\mathbf{k})$ generically contains a $\mathbf{k}$-independent term we do not expect band-touching points in the normal state.

\section{Superconducting pairing states}
\label{sec:SC_pairing_states}

In order to understand the unconventional phase diagram of CeRh\textsubscript{2}As\textsubscript{2}, we have to consider its properties that go beyond generic materials with $D_{4h}$ point group: (1) the material consists of relatively weakly coupled layers, (2) the inversion centers lie between these layers, and (3) the $\mathrm{Ce}$ \textit{f}-electrons carry a relatively large spectral weight close to the Fermi energy.

The superconducting states are described by the $8\times 8$ Bogoliubov--de Gennes (BdG) Hamiltonian
\begin{equation}
\mathcal{H}(\mathbf{k}) = \left(\begin{array}{cc}
    H_N(\mathbf{k}) & \Delta(\mathbf{k}) \\[0.5ex]
    \Delta^\dagger(\mathbf{k}) & -H_N^T(-\mathbf{k})
  \end{array}\right) .
\label{1.HBdG.2}
\end{equation}
It is useful to write the pairing matrix as
\begin{equation}
\Delta(\mathbf{k}) = D(\mathbf{k})\, U_T ,
\label{1.DeltaD.2}
\end{equation}
where $D(\mathbf{k})$, unlike $\Delta(\mathbf{k})$, transforms like a matrix under point-group transformations \cite{LiH20, timm_symmetry_2021}, which facilitates the symmetry analysis. $D(\mathbf{k})$ can be expanded into basis matrices from Table \ref{tab:irreps_4by4_basis_matrices} with momentum-dependent form factors. Unlike the normal-state block $H_N(\mathbf{k})$, $D(\mathbf{k})$ need not transform trivially ($A_{1g+}$) under the point group since the superconducting state can break the lattice symmetry spontaneously. However, contributions that are odd under TR cannot occur since only TR-even ``$+$'' irreps are compatible with fermionic antisymmetry~\cite{LiH20, timm_symmetry_2021}.

Our next task is to determine the most likely superconducting state in zero applied magnetic field. The analysis relies on two distinct ingredients: First, the pairing interaction in different symmetry channels is of course different. We will give arguments based on the specific structure of CeRh\textsubscript{2}As\textsubscript{2} as to which channels are favored. Second, even equal interaction strengths can open gaps of vastly different size at the Fermi energy, which is what matters for the weak-coupling instability. We address these two aspects in the following.

\subsection{Pairing interaction}
\label{subsec:pairing}

The specific-heat jump at the superconducting transition suggests a strongly enhanced effective mass, which implies that CeRh\textsubscript{2}As\textsubscript{2} is a heavy-fermion material with significant Ce \textit{f}-orbital weight at the Fermi energy \cite{cavanagh_nonsymmorphic_2022}. Standard density functional theory is unable to describe this mass enhancement but including renormalization of the bands due to strong correlations \cite{Zwi92, Zwi16} provides a band structure with rather flat bands with large \textit{f}-orbital weight at the Fermi energy \cite{cavanagh_nonsymmorphic_2022}. However, Nogaki \textit{et al.}\ \cite{NDI21} and Ptok \textit{et al.}\ \cite{ptok_electronic_2021} obtained rather flat \textit{f} bands without this renormalization. Because of the large \textit{f}-electron weight the onsite Hubbard repulsion $U$ is expected to be strong. Therefore, any pairing state that involves on-site pairing should be disfavored by the Hubbard repulsion. This applies to pairing matrices $\Delta(\mathbf{k}) = D(\mathbf{k})\, U_T$ that satisfy three conditions: (a) $D(\mathbf{k})$ describes pairing at the same basis site, i.e., it contains $\tau_0$ or $\tau_z$. (b) $D(\mathbf{k})$ describes spin-singlet pairing, i.e., it contains $\sigma_0$, because our model has only a single orbital per $\mathrm{Ce}$ site and local spin-triplet pairing is excluded by the Pauli principle. (c) The form factor contains a momentum-independent (local) term and thus necessarily has $A_{1g+}$ symmetry.

The basis matrices resulting from conditions (a) and (b) are \gp{0}{0} ($A_{1g+}$) and \gp{z}{0} ($A_{2u+}$). Both are compatible with an $A_{1g+}$ form factor so that condition (c) does not lead to a further reduction.

The layered nature of CeRh\textsubscript{2}As\textsubscript{2} suggests that the strongest (pairing) interactions are in plane. So far, we have argued that they are repulsive on site. Due to the localization of \textit{f}-orbitals and the relatively strong screening due to a high density of states at the Fermi energy, we do not expect strong longer-range repulsive Coulomb interactions. Hence, it is natural to assume that the dominant pairing interaction is in plane, of short range, but not local. In the absence of arguments to the contrary, the leading contribution to superconductivity is assumed to be in-plane nearest-neighbor pairing. Which pairing matrices are compatible with this?

First, they must contain $\tau_0$ or $\tau_z$ because $\tau_x$ and $\tau_y$ describe interlayer pairing. These correspond to the first and last blocks in Table \ref{tab:irreps_4by4_basis_matrices}, including irreps of $A_{1g+}$, $A_{2g-}$, $E_{g-}$, $A_{1u-}$, $A_{2u+}$, and~$E_{u-}$. Second, the form factors for in-plane nearest-neighbor pairing contain linear combinations of $\cos k_x$, $\cos k_y$, $\sin k_x$, and $\sin k_y$. These are easily organized by irreps:
\begin{align}
& \cos k_x + \cos k_y && A_{1g+}, \\
& \cos k_x - \cos k_y && B_{1g+}, \\
& (\sin k_x,\: \sin k_y) && E_{u-} .
\end{align}
The possible symmetries of in-plane nearest-neighbor pairing states are given in Table \ref{tab:possible_symmetries_in-plane_nn_pairing}. Note that all irreps of $D_{4h}$ with positive sign under TR occur in the table.

\begin{table*}[]
\caption{Possible symmetries of intralayer nearest-neighbor pairing }{states. Rows (columns) correspond to different symmetries of form factors (matrices).}
\centering
\begin{tabularx}{\linewidth}{l@{\extracolsep{\fill}}l@{~}l@{~}ll@{~}l@{~}l} \hline \hline
Form factor & $A_{1g+}$ & $A_{2g-}$ & $E_{g-}$ & $A_{1u-}$ & $A_{2u+}$ & $E_{u-}$ \\ \hline
$A_{1g+}$ & $A_{1g+}$ & & & & $A_{2u+}$ & \\
$B_{1g+}$ & $B_{1g+}$ & & & & $B_{2u+}$ & \\
$E_{u-}$  & & $E_{u+}$ & $A_{1u+} \oplus A_{2u+} \oplus B_{1u+} \oplus B_{2u+}$ &
  $E_{g+}$ & & $A_{1g+} \oplus A_{2g+} \oplus B_{1g+} \oplus B_{2g+}$ \\ \hline \hline
\end{tabularx}
\label{tab:possible_symmetries_in-plane_nn_pairing}
\end{table*}

We can now list the matrices $D(\mathbf{k})$ for all in-plane nearest-neighbor pairing states and their irreps. The spin-singlet pairing states are
\begin{align}
& (\cos k_x + \cos k_y)\, \g{0}{0} && A_{1g+} ,
\label{3.pstates.1} \\
& (\cos k_x + \cos k_y)\, \g{z}{0} && A_{2u+} , \\
& (\cos k_x - \cos k_y)\, \g{0}{0} && B_{1g+} ,
\label{3.pstates.3} \\
& (\cos k_x - \cos k_y)\, \g{z}{0} && B_{2u+} ,
\label{3.pstates.4}
\end{align}
and the spin-triplet states are
\begin{align}
& \sin(k_x)\, \g{z}{y} - \sin(k_y)\, \g{z}{x} && A_{1g+} , \label{3.pstates.5} \\
& \sin(k_x)\, \g{z}{x} + \sin(k_y)\, \g{z}{y} && A_{2g+} , \\
& \sin(k_x)\, \g{0}{x} + \sin(k_y)\, \g{0}{y} && A_{1u+} , \\
& \sin(k_x)\, \g{0}{y} - \sin(k_y)\, \g{0}{x} && A_{2u+} ,
  \\
& \sin(k_x)\, \g{z}{y} + \sin(k_y)\, \g{z}{x} && B_{1g+} , \\
& \sin(k_x)\, \g{z}{x} - \sin(k_y)\, \g{z}{y} && B_{2g+} , \label{3.pstates.7} \\
& \sin(k_x)\, \g{0}{x} - \sin(k_y)\, \g{0}{y} && B_{1u+} , \label{3.pstates.10} \\
& \sin(k_x)\, \g{0}{y} + \sin(k_y)\, \g{0}{x} && B_{2u+} , \label{3.pstates.11} \\
& (\sin k_x, \sin k_y)\, \g{z}{z} && E_{g+} , \\
& (\sin k_y, {-}\sin k_x)\, \g{0}{z} && E_{u+} , \label{3.pstates.last}
\end{align}
where the OP for the two-dimensional irreps $E_{g+}$ and $E_{u+}$ have two components. The expressions in Eqs.\ (\ref{3.pstates.5})--(\ref{3.pstates.11}) result from the reductions $E_{u-} \otimes E_{u-} = A_{1g+} \oplus A_{2g+} \oplus B_{1g+} \oplus B_{2g+}$ and $E_{u-} \otimes E_{g-} = A_{1u+} \oplus A_{2u+} \oplus B_{1u+} \oplus B_{2u+}$ together with the observation that consistently ordered doublets are $(\sin k_x,\sin k_y)$, $(\g{z}{y},-\g{z}{x})$, and $(\g{0}{x},\g{0}{y})$. In principle, $D(\mathbf{k})$ is a linear combination of these matrix-valued functions of momentum. At the critical temperature $T_c$, generically only terms belonging to the same irrep will become nonzero, though.

Note that half of the terms listed in Eqs.\ (\ref{3.pstates.1})--(\ref{3.pstates.last}) contain $\tau_0$ and thus describe superconductivity with equal pairing amplitude in neighboring $\mathrm{Ce}$ layers, while the other half contain $\tau_z$ and describe superconductivity with a sign change of the pairing amplitude between neighboring layers, which has been called a ``pair-density wave state'' in the literature \cite{YSY12, ogata_parity_2023}. Recall that the layers are weakly coupled. Weakly coupled layers can be understood as two-dimensional superconductors with Josephson junctions connecting them. BCS theory applied to generic models for a stack of two-dimensional layers with weak hybridization between them and with a local pairing interaction leads to the same pairing amplitude in all layers in equilibrium, i.e., to $0$ junctions. One has to introduce nontrivial coupling between the layers, for example, through ferromagnetic tunneling barriers or structural twisting, to favor sign changes of the pairing amplitude between layers, i.e., $\pi$ junctions. Nothing in what we know about CeRh\textsubscript{2}As\textsubscript{2} suggests that such a mechanism applies. We conclude that the pairing states containing $\tau_0$ are favored over those containing $\tau_z$.

Moreover, Eqs.\ (\ref{3.pstates.1})--(\ref{3.pstates.4}) describe spin-singlet pairing, whereas the remaining terms describe spin-triplet pairing. Observing that spin-singlet pairing is much more common than spin-triplet pairing, unless they coexist due to symmetry, it is natural to assume that the pairing interactions in the spin-singlet channels are stronger. Specifically, CeRh\textsubscript{2}As\textsubscript{2} does not show a tendency to ferromagnetic order so that we do not expect strong ferromagnetic fluctuations, which might drive triplet pairing~\cite{JCD25}\footnote{Strong AFM fluctuations may be present in CeRh\textsubscript{2}As\textsubscript{2} but these do not generally lead to triplet pairing, as seen for the cuprate and iron-pnictide superconductors.}.

The first four columns of Table \ref{tab:SC_states} summarize our conclusions concerning the bare pairing interaction in the site-spin basis. We see that $B_{1g}$ is the only pairing symmetry that avoids all restrictions: It should have a strong pairing interaction in the spin-singlet channel. There is also a spin-triplet contribution of the same symmetry \cite{YSY14}, which is disfavored since it involves sign changes of the pairing amplitude between layers, besides being a triplet state. Hence, the $B_{1g}$ spin-triplet component will not contribute much to the internal-energy gain.

\begin{table*}[]
\caption{Assessment of superconducting states restricted to in-plane nearest-neighbor pairing. The pairing states are listed in the same order as in Eqs.\ (\ref{3.pstates.1})--(\ref{3.pstates.last}). For each entry, the second to fourth columns specify whether a contribution is suppressed by local Hubbard repulsion $U$, sign change between layers ($\pi$ junctions), or spin-triplet pairing. The fifth column indicates whether the state is a pseudospin-singlet or pseudospin-triplet state. The sixth and seventh columns show the leading factor describing the rescaling of the superconducting gap at the Fermi energy, i.e., in the pseudospin basis, relative to the pairing amplitude in the original BdG Hamiltonian, for the two cases of (i) large interlayer hopping $\overline{c}_1$ and (ii) large intralayer SOC $\overline{c}_3 = \overline{c}_4$, respectively. $\overline{c}_5$ is the smallest scale in any case. For pseudospin-triplet states, the factor is distinct for the in-plane components $(d_x,d_y)$ of the $\mathbf{d}$ vector and for the out-of-plane component $d_z$.}
\centering
\begin{tabular}{lccclcc} \hline \hline
Pairing state & \multicolumn{3}{c}{Interaction} & Pseudospin state &
  \multicolumn{2}{c}{Gap renormalization} \\
(bare spin) & Hubbard $U$ & $\pi$ junction & Triplet & & Large hopping & Large SOC
  \\ \hline
$A_{1g+}$ singlet & yes & no & no & singlet & $1$ & $1$ \\
$A_{2u+}$ singlet & yes & yes & no &
  triplet $\left\{ \begin{matrix} (d_x, d_y) \\ d_z \end{matrix}\right.$ &
  $\begin{matrix} \overline{c}_3/\overline{c}_1 \\ \overline{c}_5/\overline{c}_1 \end{matrix}$ &
  $\begin{matrix} 1 \\ \overline{c}_5/\overline{c}_1 \end{matrix}$ \\
$B_{1g+}$ singlet & no & no & no & singlet & $1$ & $1$ \\
$B_{2u+}$ singlet & no & yes & no &
  triplet $\left\{ \begin{matrix} (d_x, d_y) \\ d_z \end{matrix}\right.$ &
  $\begin{matrix} \overline{c}_3/\overline{c}_1 \\ \overline{c}_5/\overline{c}_1 \end{matrix}$ &
  $\begin{matrix} 1 \\ \overline{c}_5/\overline{c}_1 \end{matrix}$ \\ \hline
$A_{1g+}$ triplet & no & yes & yes & singlet & $\overline{c}_3/\overline{c}_1$ & $1$ \\
$A_{2g+}$ triplet & no & yes & yes & singlet & $\overline{c}_3/\overline{c}_1$ & $1$ \\
$A_{1u+}$ triplet & no & no & yes &
  triplet $\left\{ \begin{matrix} (d_x, d_y) \\ d_z \end{matrix}\right.$ &
  $\begin{matrix} 1 \\ \overline{c}_3\overline{c}_5/\overline{c}_1^2 \end{matrix}$ &
  $\begin{matrix} 1 \\ \overline{c}_5/\overline{c}_1 \end{matrix}$ \\
$A_{2u+}$ triplet & no & no & yes &
  triplet $\left\{ \begin{matrix} (d_x, d_y) \\ d_z \end{matrix}\right.$ &
  $\begin{matrix} 1 \\ \overline{c}_3\overline{c}_5/\overline{c}_1^2 \end{matrix}$ &
  $\begin{matrix} 1 \\ \overline{c}_5/\overline{c}_1 \end{matrix}$ \\
$B_{1g+}$ triplet & no & yes & yes & singlet & $\overline{c}_3/\overline{c}_1$ & $1$ \\
$B_{2g+}$ triplet & no & yes & yes & singlet & $\overline{c}_3/\overline{c}_1$ & $1$ \\
$B_{1u+}$ triplet & no & no & yes &
  triplet $\left\{ \begin{matrix} (d_x, d_y) \\ d_z \end{matrix}\right.$ &
  $\begin{matrix} 1 \\ \overline{c}_3\overline{c}_5/\overline{c}_1^2 \end{matrix}$ &
  $\begin{matrix} 1 \\ \overline{c}_5/\overline{c}_1 \end{matrix}$ \\
$B_{2u+}$ triplet & no & no & yes &
  triplet $\left\{ \begin{matrix} (d_x, d_y) \\ d_z \end{matrix}\right.$ &
  $\begin{matrix} 1 \\ \overline{c}_3\overline{c}_5/\overline{c}_1^2 \end{matrix}$ &
  $\begin{matrix} 1 \\ \overline{c}_5/\overline{c}_1 \end{matrix}$ \\
$E_{g+}$ triplet & no & yes & yes & singlet & $\overline{c}_5/\overline{c}_1$ &
  $\overline{c}_5/\overline{c}_3$ \\
$E_{u+}$ triplet & no & no & yes &
  triplet $\left\{ \begin{matrix} (d_x, d_y) \\ d_z \end{matrix}\right.$ &
  $\begin{matrix} 0 \\ 1 \end{matrix}$ &
  $\begin{matrix} 0 \\ \overline{c}_1/\overline{c}_3 \end{matrix}$ \\ \hline \hline
\end{tabular}
\label{tab:SC_states}
\end{table*}

\subsection{Pseudospin picture}
\label{subsec:pseudospin}

In addition to analyzing the bare interaction, we have to assess how it acts at the Fermi energy, which will turn out to be essential for understanding the coupling of superconducting OPs to magnetic order and magnetic field. This can in principle be done by diagonalizing the BdG Hamiltonian in Eq.\ (\ref{1.HBdG.2}). A more transparent method is to first transform the BdG Hamiltonian into the pseudospin basis, which diagonalizes the normal-state blocks. Since the bands are twofold degenerate, the states can be labeled by a pseudospin of length $1/2$ \cite{Fu15, ABT17, brydon_bogoliubov_2018, cavanagh_nonsymmorphic_2022}. Then, the off-diagonal block $\Delta_{--}(\mathbf{k})$ pertaining to the low-energy band, denoted by the subscript ``$-$'', gives information on how efficient a certain superconducting state is in opening a gap. More details on the pseudospin picture are given in Appendix \ref{app:pseudospin}. BCS theory shows that a reduction of the gap at the Fermi energy by a factor of $\eta < 1$ has essentially the same effect as a reduction of the pairing interaction by a factor of $\eta^2$. The argument is sketched in Appendix \ref{app:BCSscaled}.

The projection onto the low-energy band does not change the symmetry of the superconducting state. Since the pseudospin behaves like a real spin under symmetry transformations, even-parity (odd-pa\-ri\-ty) pairing strictly corresponds to pseudospin-singlet (pseudospin-triplet) pairing. The pairing matrix can be written as $\Delta_{--}(\mathbf{k}) = \psi(\mathbf{k})\, is_y$ for pseudospin-singlet pairing and as $\Delta_{--}(\mathbf{k}) = \mathbf{d}(\mathbf{k})\cdot \mathbf{s}\, is_y$ for pseudospin-triplet pairing \cite{sigrist_phenomenological_1991}, where $\mathbf{s} = (s_x, s_y, s_z)$ are the Pauli matrices on pseudospin space. In the following, when we mention the spin---as opposed to the pseudospin---we mean the bare spin described by Pauli matrices $\sigma_x$, $\sigma_y$, and $\sigma_z$.

The pairing amplitude in the original BdG Hamiltonian, Eq.\ (\ref{1.HBdG.2}), is rescaled by a factor that depends on the pairing state, i.e., on $\Delta(\mathbf{k})$. For any pairing state, in particular, for any entry in Table \ref{tab:SC_states}, the factor can be expressed in terms of the form factors $c_n(\mathbf{k})$, $n = 1,\ldots,5$, appearing in the normal-state Hamiltonian $H_N(\mathbf{k})$. It is therefore crucial to understand the hierarchy of energy scales encoded in $c_n(\mathbf{k})$. We define the typical energy of the five terms by
\begin{equation}
\overline{c}_n = \sqrt{ \langle c_n^2(\mathbf{k}) \rangle_\mathrm{FS} } ,
\end{equation}
where the average is over the Fermi surface. The physical interpretation of the five terms is given in Sec.\ \ref{sec:model}, following Eq.\ (\ref{1.basis.c5}). As noted there, the long-range interlayer SOC $\overline{c}_5$ is expected to be small compared to both the interlayer hoppings $\overline{c}_1$ and $\overline{c}_2$ and the nearest-neighbor intralayer SOC $\overline{c}_3 = \overline{c}_4$ \footnote{The averages $\overline{c}_3$ and $\overline{c}_4$ must be equal because $c_3(\mathbf{k})$ and $c_4(\mathbf{k})$ are basis functions of the two components of $E_u$.}. The relative size of $\overline{c}_1$, $\overline{c}_2$ vs.\ $\overline{c}_3 = \overline{c}_4$ is less clear. The band structure obtained from ab-initio calculations shows sizable dispersion in the \textit{z}-direction \cite{NDI21, hafner_possible_2022, cavanagh_nonsymmorphic_2022}, suggesting that $\overline{c}_1$ and perhaps $\overline{c}_2$ are not small. On the other hand, CeRh\textsubscript{2}As\textsubscript{2} contains only heavy elements with relatively strong SOC. Cavanagh \textit{et al.}\ \cite{cavanagh_nonsymmorphic_2022} show that the nonsymmorphic space group leads to SOC being larger than interlayer hopping at the boundary of the Brillouin zone, where significant portions of the Fermi surface seem to reside.

We here consider the extremal cases of (i) large interlayer hopping, $\overline{c}_1 \gg \overline{c}_3 = \overline{c}_4 \gg \overline{c}_5$ ($\overline{c}_2$ turns out not to matter in this case), and (ii) large SOC, $\overline{c}_3 = \overline{c}_4 \gg \overline{c}_1 \approx \overline{c}_2 \gg \overline{c}_5$. The leading factors describing the rescaling of the superconducting gap at the Fermi energy, i.e., in the pseudospin basis, relative to the pairing amplitude in the original BdG Hamiltonian are given in the last two columns in Table \ref{tab:SC_states}. For the pseudospin-triplet states, the factor is different for the in-plane and out-of-plane components of the $\mathbf{d}$ vector. The stability of a pseudospin-triplet state is determined by the pairing amplitude $|\mathbf{d}(\mathbf{k})|$ and is thus controlled by the largest component of the $\mathbf{d}$ vector.

The most important result is that the $B_{1g+}$ pairing state preferred by the bare interaction is not affected by the projection into the low-energy sector. The gap at the Fermi energy is not suppressed for either limiting case. Next, we note that the---in any case not very promising---$E_{g+}$ pairing state is highly disfavored since the gap at the Fermi energy contains the small factor $c_5(\mathbf{k})$. For the $E_{u+}$ state, the pseudospin $\mathbf{d}$ vector is strictly out of plane, whereas for all other pseudospin-triplet states, it is predominantly in plane~\cite{cavanagh_nonsymmorphic_2022}.

Further results depend on the limit (i) or (ii). We discuss one point: From the beginning, it was argued that the superconductor-to-superconductor transition in a magnetic field of about $4\,\mathrm{T}$ along the \textit{z}-direction indicated the existence of two nearly degenerate superconducting states. It was suggested that the weak coupling between layers implied that superconducting states that differ only in the relative sign of the pairing amplitude in neighboring layers are close in energy \cite{khim_field-induced_2021, cavanagh_nonsymmorphic_2022}. The corresponding pairs of states can easily be read off from Eqs.\ (\ref{3.pstates.1})--(\ref{3.pstates.last}). We call them ``partner states.'' In particular, the $B_{1g+}$ spin-singlet state and the $B_{2u+}$ spin-singlet state are partner states. Hence, it seems natural to assume that the low-field phase is dominated by $B_{1g+}$ and the high-field phase by $B_{2u+}$.

We want to make the point that the conclusion of near degeneracy of $B_{1g+}$ and $B_{2u+}$ states is not generally true but strongly depends on the hierarchy of normal-state energy scales $\overline{c}_n$. First, it is indeed natural to assume that the bare interactions for partner states are similar (the effect of the ``yes'' or ``no'' in the column headed ``$\pi$ junctions'' in Table \ref{tab:SC_states} is small). Second, the renormalization of the gap at the Fermi energy can be a strong reduction. Such a reduction leads to an even stronger (quadratic) suppression of the effective pairing interaction. Thanks to the weak-coupling expression for the gap in BCS theory, this causes an exponential suppression of the selfconsistent pairing amplitude at zero temperature and also of the critical temperature, see Appendix \ref{app:BCSscaled}. Now, consider the purported partner states $B_{1g+}$ (spin-singlet) and $B_{2u+}$ (spin-singlet) in Table \ref{tab:SC_states}. The $B_{1g+}$ state is unaffected by the renormalization. If the interlayer hopping is large compared to the intralayer SOC, then the renormalization factor for the relevant in-plane component of the $B_{2u+}$ state is on the order of $\overline{c}_3/\overline{c}_1 \ll 1$. The effective pairing interaction is then reduced by $\overline{c}_3^2/\overline{c}_1^2 \ll 1$ and the selfconsistent gap for $B_{2u+}$ is exponentially smaller than for $B_{1g+}$. On the other hand, if the intralayer SOC is the largest scale the renormalization factor for $B_{2u+}$ is on the order of unity. In this limit, we obtain similar zero-temperature selfconsistent gaps and critical temperatures in the $B_{1g+}$ and $B_{2u+}$ spin-singlet channels, in agreement with Ref.\ \cite{cavanagh_nonsymmorphic_2022}.

We can now turn the argument around: Evidently, there are two superconducting states with similar energies. The only way to avoid a fine-tuning problem is to conclude that the intralayer SOC is larger than the interlayer hopping. An alternative mechanism has been proposed by Nally and Brydon \cite{NaB24}, who apply a strong-coupling theory to a \textit{t}--\textit{J} model: Superconducting states of certain symmetries are disfavored if their gap nodes lie close to large parts of the Fermi surface. In such a case, the typical gap at the Fermi energy, which drives the weak-coupling instability, is suppressed compared to the naive maximum gap. In Ref.\ \cite{NaB24}, a transition between \textit{A}-type (extended \textit{s}-wave) and \textit{B}-type (\textit{d}-wave) states is tuned by shifting the Fermi surface through the gap nodes by changing the strength of spin-orbit coupling.

Based on the above reasoning, the most likely pairing state at weak magnetic fields and ignoring magnetic order has $B_{1g+}$ symmetry. The superconducting OP changes sign in momentum space, unlike for $A_{1g+}$ pairing. Sign-changing superconducting states are expected to be sensitive to disorder, which is consistent with the observed variation of critical temperatures $T_c$ with sample quality \cite{khanenko_phase_2025}. On the other hand, the first-order superconductor-to-superconductor transition is quite robust. The effect of disorder has been studied theoretically by Cavanagh and Brydon \cite{CaB21} and by M\"ockli and Ramires \cite{MoR21}. The upshot is that the robustness of a superconducting OP of given symmetry against disorder depends strongly on the symmetry, in sublattice and spin space, of the impurities. However, the observation that the first-order transition varies little between samples indicates that such a mechanism is not dominant for CeRh\textsubscript{2}As\textsubscript{2}.

Our result is consistent with the $d_{x^2-y^2}$-pairing state recently proposed by Amin \textit{et al.}\ \cite{amin_kramers_2024}. In that work, Kramers' degenerate magnetic order, i.e., magnetic order that breaks inversion and TR symmetry but preserves their product, and its interplay with superconductivity are considered. We address the magnetic order in the following section. The closely competing $B_{1g}$ and $B_{2u}$ states are also found by Lee \textit{et al.}\ \cite{lee_unified_2024} from a renormalization-group analysis of a $\mathbf{k}\cdot\mathbf{p}$ model centered at the X point. However, Lee \textit{et al.}\ \cite{lee_unified_2024} find that the $B_{2u}$ state is stabilized in large parameter regions. Pairing states that are insensitive to the local interaction have also been found by Nogaki and Yanase \cite{nogaki_even-odd_2022} using the fluctuation exchange approximation, by Nally and Brydon \cite{NaB24}, and by Lee \textit{et al.}\ \cite{lee_unified_2024} with a parquet renormalization-group approach.

\section{Magnetic ordering}
\label{sec:magnetism}

Magnetic ordering in CeRh\textsubscript{2}As\textsubscript{2} was first inferred from NQR and NMR experiments \cite{kibune_observation_2022, ogata_parity_2023}. Here, the onset of magnetism was found to occur below the superconducting transition, i.e., at a N\'eel temperature $T_N < T_c$. These measurements are most suited to identify the symmetry of the magnetic order, as discussed in the following subsection. Phase I mentioned in Sec.\ \ref{sec:intro} was interpreted in terms of quadrupolar order \cite{hafner_possible_2022, mishra_anisotropic_2022}. Recently, this scenario has been challenged by $\mu$SR studies; Khim \textit{et al.}\ \cite{khim_coexistence_2024} suggest a magnetic dipolar character of phase~I.

\begin{figure*}[t]
\centering
\includegraphics[width=.8\textwidth]{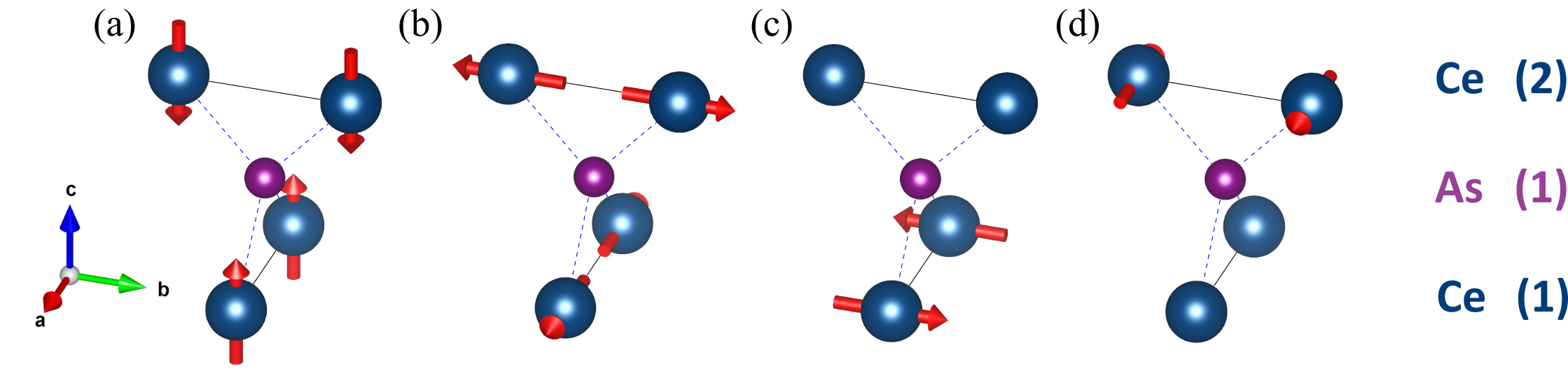}
\caption{The four linearly independent spin configurations at the four Ce neighbors of an As(1) site.}
\label{fig:As1neighbors}
\end{figure*}

\begin{figure*}[t]
\centering
\includegraphics[width=.9\textwidth]{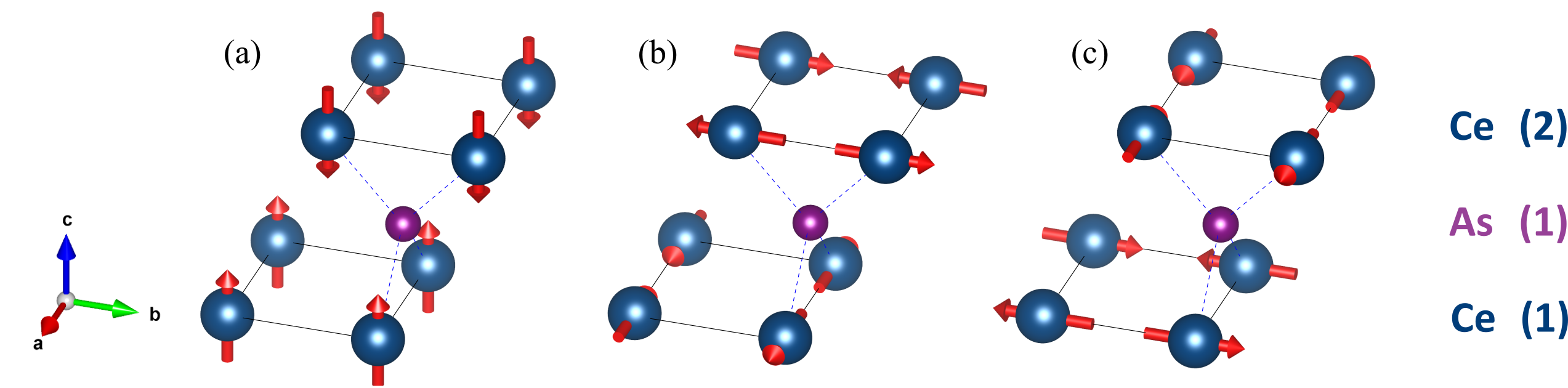}
\caption{The three linearly independent translationally invariant magnetic orders of a Ce double layer consistent with the absence of a magnetic field at the $\mathrm{As}(1)$ sites. Configurations (b) and (c) are degenerate. All linear combinations are also compatible with the zero-field condition.}
\label{fig:magnetic_order_As1_zero_field}
\end{figure*}

\subsection{Magnetic order parameters from NQR and NMR}
\label{subsec:M_from_NMR_NQR}

In this section, we discuss the possible magnetic OPs in view of the NQR experiments by Kibune \textit{et al.}\ \cite{kibune_observation_2022} and NMR experiments by Ogata \textit{et al.}\ \cite{ogata_parity_2023, ogata_appearance_2024}. The NQR and NMR experiments are sensitive to the local magnetic field at $\mathrm{As}$ positions. There are two inequivalent positions, see Fig.~\ref{fig:level_splitting}: $\mathrm{As}(1)$ forms a square lattice in the middle between two $\mathrm{Ce}$ layers. The $\mathrm{As}(1)$ square lattice is rotated by $45^\circ$ relative to the $\mathrm{Ce}$ square lattice and contains twice as many sites. $\mathrm{As}(2)$ is vertically aligned with $\mathrm{Ce}$ sites and does not lie in the middle between $\mathrm{Ce}$ layers. The central result of Ref.\ \cite{kibune_observation_2022} is that in the absence of an applied magnetic field, the local magnetic field at the $\mathrm{As}(1)$ site is consistent with zero, whereas the local field at $\mathrm{As}(2)$ is clearly nonzero, as shown by a strong broadening of the NQR peak. The NMR experiments \cite{ogata_parity_2023} are consistent with this result.

It is natural to attribute the magnetic order to the $\mathrm{Ce}$ \textit{f} moments \cite{ptok_electronic_2021}. Alternatively, it may be the consequence of TR-symmetry-breaking superconductivity, which is carried by electrons at the Fermi energy, which also have a high $\mathrm{Ce}$ \textit{f}-orbital weight. In the following, we determine the spin configurations that are consistent with vanishing local magnetic field at all As(1) sites and nonzero field, of equal magnitude, at all As(2) sites, in view of the NQR experiments \cite{kibune_observation_2022}, without prior assumptions on translational invariance.

For every As(1) site, at least the field due to the nearest four Ce sites, which are related by $S_{4z}$ transformations, must cancel. The spin configuration of these four sites must therefore be a linear combination of the four configurations shown in Fig.\ \ref{fig:As1neighbors}. The requirement that the local fields at \emph{all} As(1) sites vanish then implies that the orientation of one Ce spin fixes all the others in the same double layer. A double layer is here defined as two Ce layers with As(1) sites between them. Hence, the spin configuration of every double layer is translationally invariant and is a linear combination of the three spin orders shown in Fig.~\ref{fig:magnetic_order_As1_zero_field}.

This leaves us to determine the relative spin orientation in different double layers. Double layers separated by $2c$, where $c$ is the lattice constant in the \textit{z}-direction, contain quartets of Ce sites related by the same $S_{4z}$ symmetry centered at an As(1) site as above \footnote{This constraint relies on longer-range dipole fields. The field at the $\mathrm{As}(1)$ sites from breaking these constraints may be too weak to observe.}. The zero-field condition then forces double layers separated by $2c$ and hence by an even multiple of $c$ to have the same spin configuration.

To determine the relative spin orientation in even-numbered and odd-numbered double layers, we turn to the As(2) sites. Each As(2) site is close to the center of a square of Ce and slightly displaced in the \textit{z}-direction. For the magnitude of the magnetic field at all As(2) sites to be equal, the spin configurations of all elementary Ce squares must be related by symmetry transformations that leave the field magnitude at the As(2) site invariant. The possible transformations are generated by time reversal $\mathcal{T}$, which flips all spins, and $\pi/2$ rotations $C_{4z}$ of all spins. Within the same double layer and for double layers separated by an even multiple of $c$, this condition does not constrain the spin configuration any further since it is satisfied by the three configurations in Fig.~\ref{fig:magnetic_order_As1_zero_field}.

\begin{figure*}[t]
\centering
\includegraphics[width=.9\textwidth]{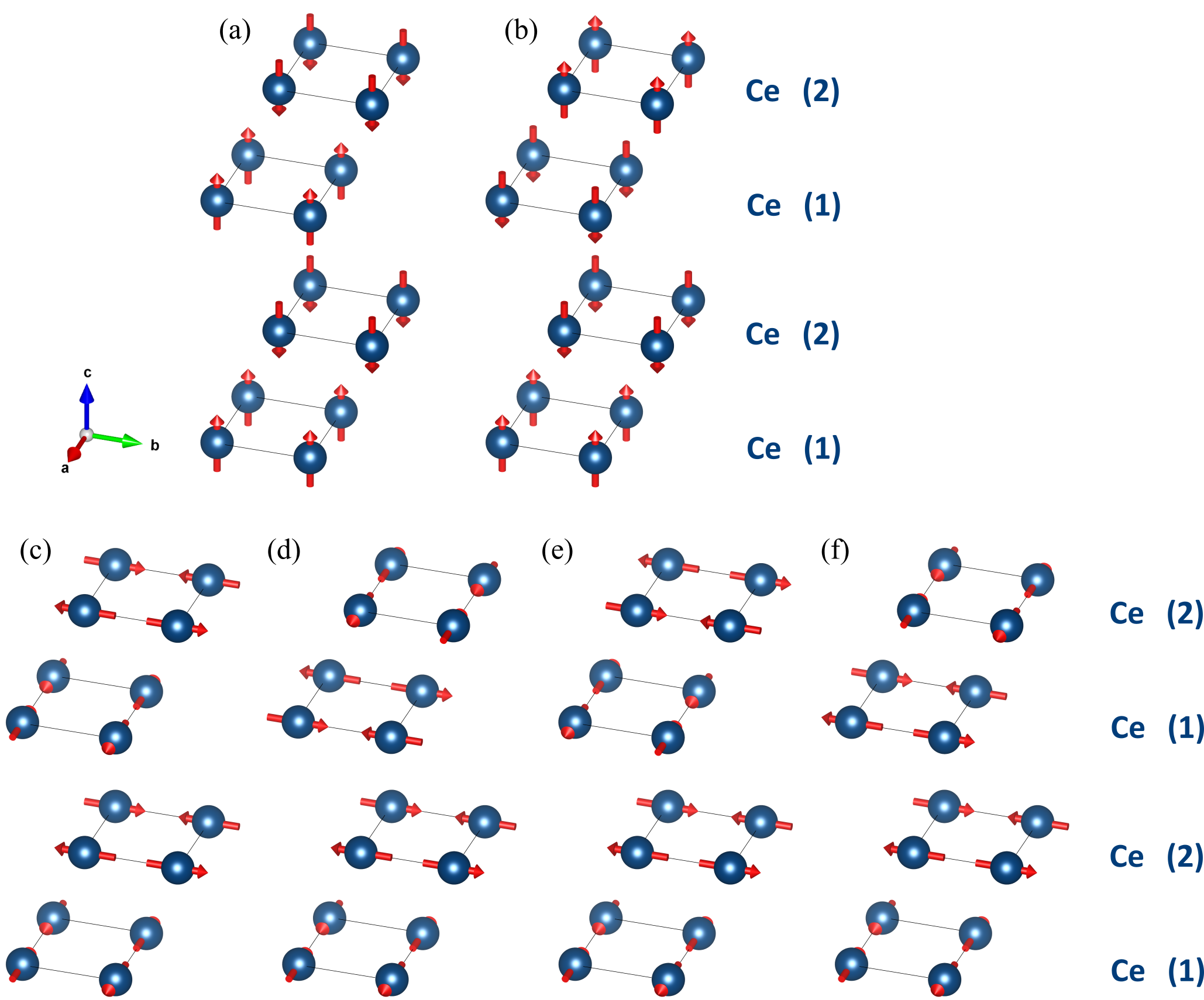}
\caption{The six linearly independent magnetic orders consistent with vanishing local magnetic field at the As(1) sites and nonvanishing but equal field magnitude at the As(2) sites.}
\label{fig:6orders}
\end{figure*}

For double layers separated by an odd multiple of $c$, we consider the out-of-plane and in-plane configurations separately, which is allowed due to linearity \footnote{The relative spin orientation in double layers separated by an odd multiple of $c$ could, in principle, be constrained by measuring the magnetic field at the $\mathrm{Rh}(1)$ sites.}. The out-of-plane spin components can either be the same or time reversed (opposite). For the in-plane components, four distinct configurations are possible. The two double layers may either share the same spin arrangement or differ by a relative rotation of all spins. This relative rotation can be generated by rotating the spins in one single layer by $\pi/2$, $\pi$, or $3\pi/2$, after which the spin orientation of the second single layer is determined by the vanishing-field condition at the $\mathrm{As}(1)$ sites. Time reversal is redundant here. Hence, the spin configuration at all Ce sites is a linear combination of six possible orders, which are shown in Fig.~\ref{fig:6orders}.

All orders are commensurate. Only the order in Fig.\ \ref{fig:6orders}(a) has ordering vector $\mathbf{q} = 0$.  Two of them are collinear out-of-plane configurations and four are noncollinear in-plane configurations. The latter involve rotations of spins by multiples of $\pi/2$ between adjacent Ce layers, which seem unnatural since there is no reason for the interlayer exchange interaction to favor such ordering---the exchange tensor would have to be strongly off-diagonal. One of the out-of-plane configurations and two of the in-plane configurations require very different exchange interactions between Ce layers separated by Rh(2)-As(1)-Rh(2) and by As(2)-Rh(1)-As(2). Again, there is no plausible mechanism for this.

Also, only the $\mathbf{q} = 0$ order in Fig.\ \ref{fig:6orders}(a) corresponds to a pure irrep, namely, $A_{1u-}$, whereas the other five are neither even nor odd under all inversions and thus mix even and odd irreps. A pure irrep is preferred because mixed irreps generically lead to separate phase transitions where the OPs of different symmetries set in, for which there is no evidence for CeRh\textsubscript{2}As\textsubscript{2}. Note that if one allows for translational symmetry to be broken, the nonsymmorphic space group $P4/nmm$ leads to the appearance of irreps that mix even and odd parity, as recently studied by Szab\'o and Ramires \cite{szabo_superconductivity-induced_2024} for the simplest case that the unit cell is doubled.

\begin{figure*}[t]
\centering
\includegraphics[width=.9\textwidth]{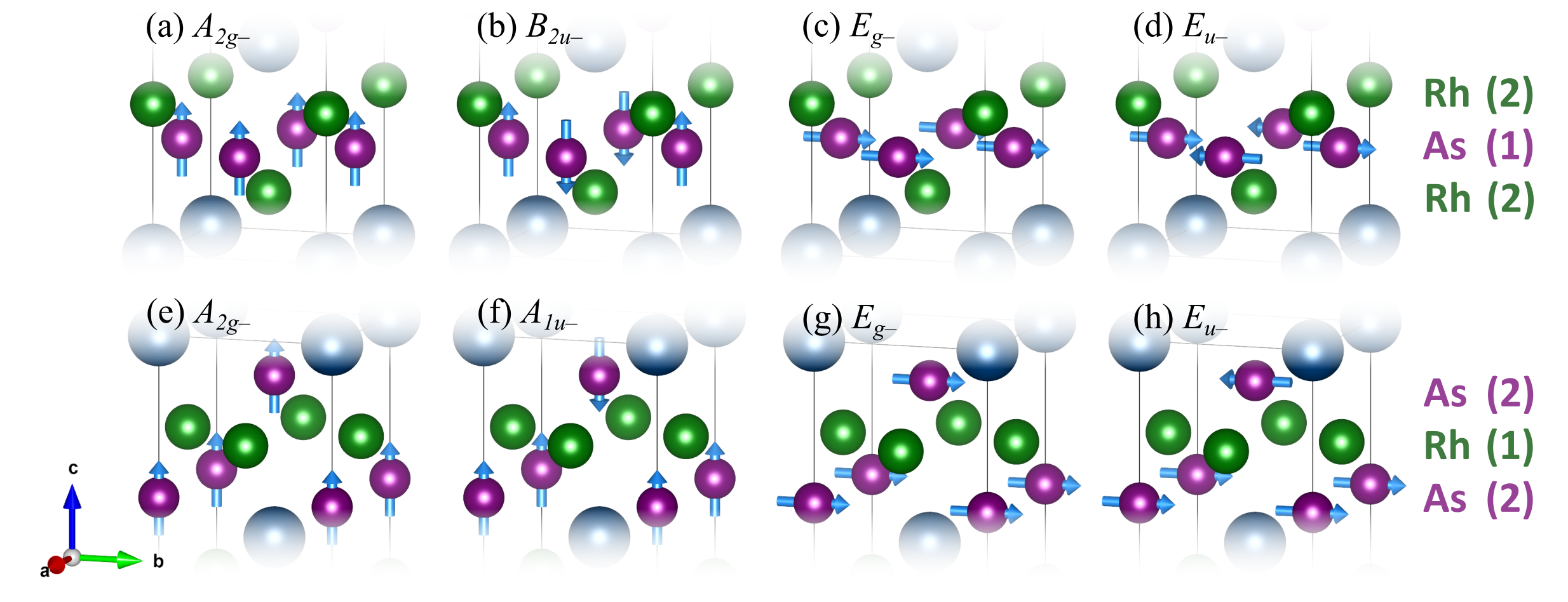}
\caption{Irreducible representations of possible magnetic orders for (a)--(d) the $\mathrm{As}(1)$ and (e)--(h) the $\mathrm{As}(2)$ sites: (a), (e) out-of-plane FM, (b), (f) out-of-plane AFM, (c), (g) in-plane FM, and (d), (h) in-plane AFM. Blue arrows describe the local magnetic field. The corresponding irreps are given in all panels.}
\label{fig:magnetic_order_As1_As2}
\end{figure*}

Next, we analyze the symmetry of possible magnetic states that do not break translational symmetry. The magnetic field at the $\mathrm{As}(1)$ and $\mathrm{As}(2)$ sites can be decomposed into components along the lattice axes. These components can be classified according to irreps, by checking the action of the operations from $D_{4h}$ on the local fields. Of course, the local field could, in principle, point in any direction but if it has in-plane and \textit{z}-axis components this would imply a state with mixed irreps, which is not expected generically. The irreps for the $\mathrm{As}(1)$ and $\mathrm{As}(2)$ sites are shown in Fig.~\ref{fig:magnetic_order_As1_As2}.

\begin{figure}[b]
\centering
\includegraphics[width=.35\textwidth]{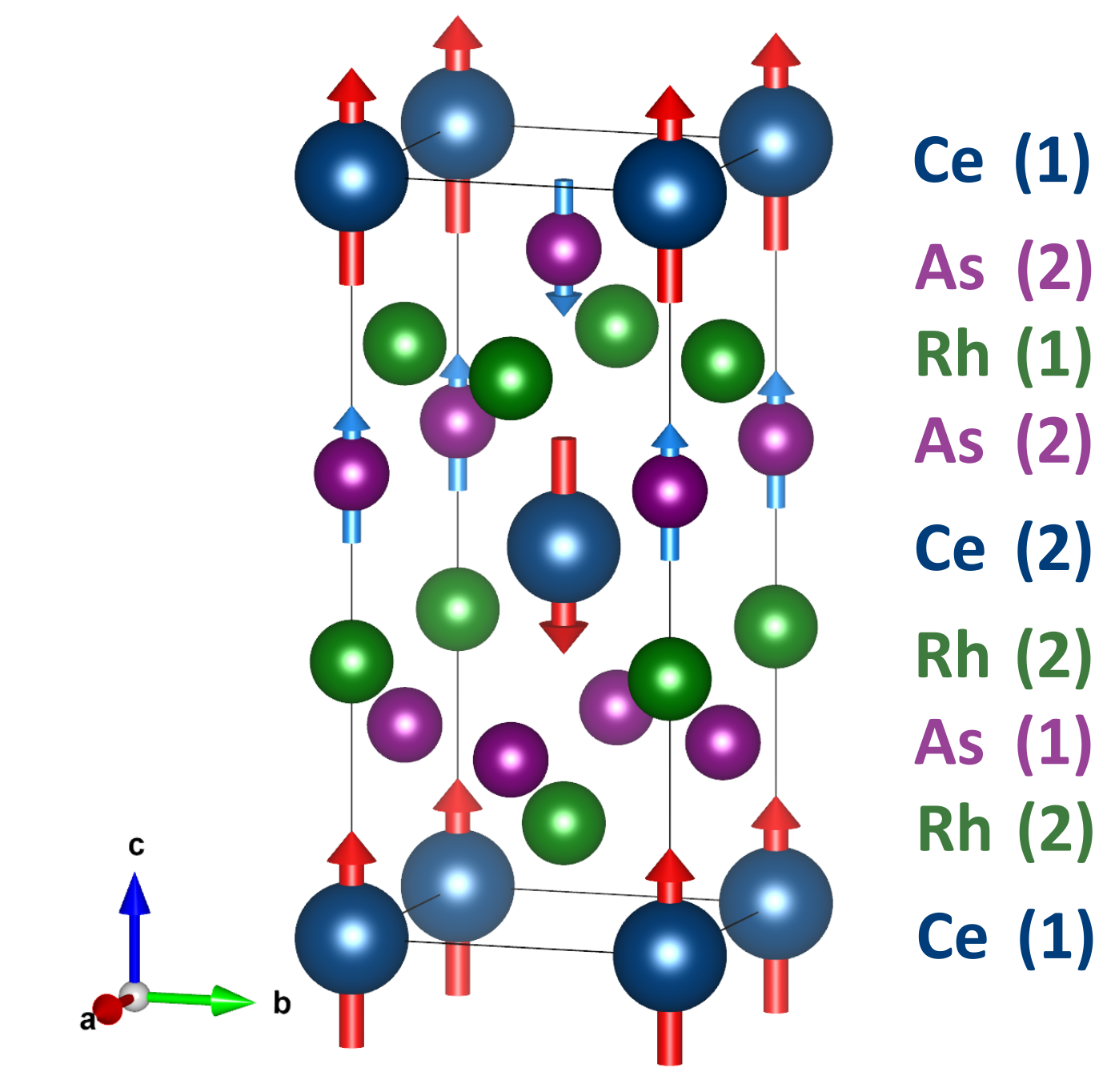}
\caption{Proposed out-of-plane AFM $A_{1u-}$ order of CeRh\textsubscript{2}As\textsubscript{2} with zero field at $\mathrm{As}(1)$ and nonzero field at $\mathrm{As}(2)$ sites.
Large orange (small blue) arrows describe the magnetic moments at Ce sites (local magnetic field at As sites).}
\label{fig:A1u_magnetic_order}
\end{figure}

We see that the only magnetic symmetry consistent with zero field at $\mathrm{As}(1)$ and nonzero field at $\mathrm{As}(2)$ is $A_{1u-}$, i.e., an out-of-plane AFM order \footnote{Conversely, $A_{2g-}$, $E_{g-}$, and $E_{u-}$ would lead to nonzero fields at all $\mathrm{As}$ sites, $B_{2u-}$ to nonzero field only at $\mathrm{As}(1)$ sites, and all other TR-odd irreps to vanishing field at all $\mathrm{As}$ sites.}. This proposed $A_{1u-}$ order is shown in Fig.~\ref{fig:A1u_magnetic_order}. Szab\'o and Ramires \cite{szabo_superconductivity-induced_2024} have come to the same conclusion. Note that $A_{1u-}$ magnetic order is odd under inversion and under TR but even under their product. This is the case considered by Amin \textit{et al.}\ \cite{amin_kramers_2024}. On the other hand, the out-of-plane N\'eel vector of the $A_{1u-}$ order seems to contradict the experimentally favored easy plane anisotropy \cite{kitagawa_two-dimensional_2022}. However, new NMR results \cite{ogata_appearance_2024} indicate that the moment is indeed out of plane, consistent with $A_{1u-}$ symmetry.

As noted above, the magnetic order is likely predominantly carried by the $\mathrm{Ce}$ atoms. The symmetry classification is analogous to the case of the $\mathrm{As}(2)$ sites. The result is again that the only pure irrep consistent with the NQR results is $A_{1u-}$, see the left sketch in {Fig.~3(a)} in Ref.\ \cite{kibune_observation_2022}. Note that the right sketch shows magnetic order that breaks translational symmetry. It is neither even nor odd under inversion at \emph{any} center and is thus not a pure-irrep order.

More recent NMR experiments \cite{ogata_parity_2023} shed additional light on the magnetic state. The authors still find a transition to a TR-symmetry-breaking state in zero field, at a temperature below the superconducting $T_c$ \cite{kibune_observation_2022}. In addition, this magnetic transition is now followed to nonzero magnetic fields along the \textit{z}-direction \cite{ogata_parity_2023}. Its critical temperature decreases with increasing $B_z$ but slightly less so than the superconducting $T_c$. The NMR results are consistent with the magnetic transition line ending where it reaches the first-order superconductor-to-superconductor transition at about $4\,\mathrm{T}$, which is the interpretation given by the authors in Ref.\ \cite{ogata_parity_2023}.

As noted above, recent $\mu$SR experiments \cite{khim_coexistence_2024} suggest that phase I in the normal state has magnetic dipolar character, i.e., that $T_N > T_c$. The internal magnetic field seen by $\mu$SR is continuous through $T_c$, hence the magnetic order is likely of the same type above and below $T_c$. Moreover, the internal field is found to be roughly temperature independent below $T_c$ instead of rising further upon cooling \cite{khim_coexistence_2024}. This indicates that magnetic order and superconductivity are coupled and compete with each other \cite{khim_coexistence_2024}. Recently, the magnetic transition could be observed within the high-field superconducting phase, where the critical field is also roughly temperature independent \cite{khanenko_phase_2025}. The origin of the distinct magnetic ordering temperatures might be different quality of older and newer samples or that the magnetic order is not static. In such a situation, the different timescales of the experimental probes lead to distinct critical temperatures \cite{khim_coexistence_2024}. We consider both the cases $T_N > T_c$ and $T_N < T_c$ in our analysis.

Note that the application of a uniform magnetic field leads to a FM contribution in addition to the out-of-plane AFM $A_{1u-}$ order. Such a field can be decomposed into a \textit{z}-component $B_z$, which transforms according to $A_{2g-}$, and in-plane components $(B_x,B_y)$, which transform according to $E_{g-}$. Thus, the expected FM OP symmetries are $A_{2g-}$ and $E_{g-}$, respectively. One could speak of a canted AFM state in this case. Compared to the pure $A_{1u-}$ state, this would lead to an increased line width also at the $\mathrm{As}(1)$ sites for finite fields. The comparison of out-of-plane NMR measurements \cite{ogata_parity_2023} with recent in-plane results \cite{ogata_appearance_2024} suggests that this effect could be especially pronounced for fields within the plane, even at low field amplitudes. This could contribute to the explanation for why a significant site-dependent broadening is observed for $\mathbf{B} \parallel [001]$ \cite{ogata_parity_2023} but not for $\mathbf{B} \parallel [110]$ \cite{ogata_appearance_2024}. However, in this work we focus on the primary AFM $A_{1u-}$ OP, which we expect to be the most relevant contribution for the purpose of understanding the superconducting phase transitions.

Results of neutron scattering experiments \cite{CSX24} are also relevant here. Elastic neutron scattering places an upper limit of $0.31\,\mu_B$ on the local moments at the Ce sites for static magnetic order with ordering vector $\mathbf{q} \neq 0$. On the other hand, inelastic neutron scattering shows quasi-two-dimensional spin fluctuations peaked at $(q_x,q_y) = (\pi/a,\pi/a)$. The authors of Ref. \cite{CSX24} interpret these spin fluctuations in terms of Fermi-surface nesting and suggest that they play a role in the superconducting pairing interaction. The measurements \cite{CSX24} are not sensitive at $\mathbf{q} = 0$. One can thus conclude that any static magnetic order, in particular, the one suggested to exist in phase I, is either weak or has ordering vector $\mathbf{q} = 0$, which is compatible with our conclusions.

\subsection{Coexistence of superconducting and magnetic order parameters}
\label{subsec:coex_SC_M}

If superconductivity and magnetic order coexist for zero applied field and the two OPs do not belong to the same irrep, then a secondary superconducting OP is induced. In the language of Landau theory, there are allowed trilinear terms of the form~\cite{szabo_superconductivity-induced_2024}
\begin{equation}
M\, i\, (\Delta_1^* \Delta_2 - \Delta_2^* \Delta_1) ,
\label{eq:trilinear_coupling}
\end{equation}
where $M$ is the magnetic OP, $\Delta_1$ and $\Delta_2$ are superconducting OPs, and the term must have $A_{1g+}$, i.e., full, symmetry. The trilinear term is reminiscent of a so-called Lifshitz invariant \cite{SSY17}. We denote the irreps of the three OPs by $\Gamma_M$, $\Gamma_1$, and $\Gamma_2$. Since the likely primary superconducting OP and the magnetic OP were found to belong to one-dimensional irreps, we restrict ourselves to this situation. It is then easy to see that the irrep of the induced OP is also one dimensional. Any symmetry-allowed term must then satisfy
\begin{equation}
A_{1g+} = \Gamma_M \otimes A_{1g-} \otimes \Gamma_1
  \otimes \Gamma_2 ,
\end{equation}
where the additional factor $A_{1g-}$ (odd under TR, otherwise trivial) appears because $\Delta_1^* \Delta_2 - \Delta_2^* \Delta_1$ is odd under TR since it maps to $\Delta_1 \Delta_2^* - \Delta_2 \Delta_1^* = -(\Delta_1^* \Delta_2 - \Delta_2^* \Delta_1)$. Hence, the induced superconducting OP belongs to the irrep
\begin{equation}
\Gamma_2 = A_{1g-} \otimes \Gamma_M \otimes \Gamma_1 .
\end{equation}
As discussed above, we propose $\Gamma_M = A_{1u-}$ and $\Gamma_1 = B_{1g+}$ at vanishing magnetic field. Hence, we obtain $\Gamma_2 = B_{1u+}$ for the induced OP. The partner state $B_{2u+}$ of $B_{1g+}$ would give $\Gamma_1 = B_{2u+}$ and $\Gamma_2 = B_{2g+}$. The argument so far is based on Landau theory. We note that in the microscopic pseudospin picture, coexisting $B_{1g+}$ and $B_{1u+}$ as well as coexisting $B_{2u+}$ and $B_{2g+}$ OPs indeed generate trilinear couplings to $A_{1u-}$ magnetic order. In both cases, this coupling appears as a correction $\gamma(\mathbf{k})$ of $A_{1u-}$ symmetry to the normal-state dispersion due to interband pairing, see Appendix~\ref{app:pseudospin}.

\section{Field-induced superconducting orders and pair breaking}
\label{sec:field_induced}

In this section, we investigate the effects of applying a magnetic field in the presence of superconducting order, where the field generically induces further superconducting OPs. In the bulk, this only happens if the field actually penetrates the sample, i.e., in the vortex phase. The symmetry of the induced OP can be determined in the standard way using product representations. This mechanism can stabilize superconducting states, as discussed below. Moreover, the magnetic field can suppress and eventually destroy superconducting condensates, in particular, due to Pauli limiting, which is discussed in Sec.~\ref{subsec:Pauli_limiting}.

\subsection{Out-of-plane field}
\label{subsec:coupling_out-of-plane_fields}

We first consider the case of a field in the \textit{z}-direction, which transforms according to $A_{2g-}$. In Landau theory, and assuming that the irreps of the superconducting states are one dimensional, the induction of another OP is described by a symmetry-allowed trilinear term
\begin{equation}
F_3 = B_z\, i\, (\Delta_1^* \Delta_2 - \Delta_2^* \Delta_1) .
\label{eq:BzD1D2}
\end{equation}
This coupling generally contains contributions both from charge currents carried by the condensate and from local spin and orbital degrees of freedom \cite{WJB25}. The irrep of the induced superconducting OP is
\begin{equation}
\Gamma_2 = A_{2g-} \otimes A_{1g-} \otimes \Gamma_1
  = A_{2g+} \otimes \Gamma_1 ,
\label{eq:G2A2gG1}
\end{equation}
where the factor $A_{1g-}$ appears because $\Delta_1^* \Delta_2 - \Delta_2^* \Delta_1$ is odd under TR, see Sec.\ \ref{subsec:coex_SC_M}. Hence, for $\Gamma_1 = B_{1g+}$, the field induces $B_{2g+}$ pairing. When two superconducting OPs of symmetries $B_{1g+}$ and $B_{1u+}$ coexist, as argued in Sec.\ \ref{subsec:coex_SC_M}, the applied field induces two additional OPs $B_{2g+}$ and $B_{2u+}$.

One of the most striking properties of CeRh\textsubscript{2}As\textsubscript{2} is the first-order superconductor-to-superconductor transition at an applied magnetic field of about $4\,\mathrm{T}$. This field is likely much larger than the lower critical field \cite{ogata_parity_2023}. Hence, the magnetic field penetrates in the form of overlapping vortices and the energy cost of magnetic-field expulsion is strongly reduced compared to the Meissner state. In the high-field phase, we again get a pair (or possibly two pairs) of superconducting OPs related by Eq.~(\ref{eq:G2A2gG1}).

For weak applied field, the induced OPs will be linear in $B_z$ and hence small. The trilinear coupling $F_3$ from Eq.\ (\ref{eq:BzD1D2}) then lowers the free energy quadratically in $B_z$. This mechanism increases the stability of superconductivity in an applied field $B_z$.

We now consider the relevant OPs in more detail. For the $B_{1g+}$ OP, a $B_{2g+}$ OP is induced. The only relevant contribution to it is an unfavorable spin-triplet state with $\pi$ junctions between layers, given in Eq.\ (\ref{3.pstates.7}). On the other hand, for the $B_{1u+}$ OP, a $B_{2u+}$ OP is induced, which has been argued to be only slightly higher in energy than $B_{1g+}$. Moreover, the $B_{2u+}$ OP involves a spin-triplet state with energetically favored $0$ junctions between layers, Eq.\ (\ref{3.pstates.11}). Therefore, the induced OPs shift the balance from the combination of $B_{1g+}$ and $B_{2g+}$ toward the combination of $B_{1u+}$ and $B_{2u+}$. It is, however, unlikely that this mechanism alone can stabilize the second combination.

For the coupling of superconducting OPs to the magnetic field, only effects at the energy scale of the superconducting gap are relevant. In other words, the bare-spin content of the pairing state is irrelevant for the coupling, only the pseudospin content matters. Hence, the pseudospin picture is again very useful. It shows that the coupling of the field $B_z$ to the odd-parity pairing states is parametrically stronger than the coupling to the even-parity states. This is because for the combination of $B_{2u+}$ and $B_{1u+}$ OPs, the effect already appears at order zero in the interband pairing: The quasiparticle dispersion is already split by intraband pairing, as can be shown by evaluating the intraband gap product $\Delta_{--} \Delta_{--}^\dagger$, see Appendix \ref{app:pseudospin}. On the other hand, the combination of $B_{1g+}$ and $B_{2g+}$ OPs affects the dispersion only perturbatively through the pseudomagnetic field induced by interband pairing. The latter contribution is generically suppressed compared to the former by a factor of the pairing amplitude divided by the interband splitting. Hence, the prefactor of $F_3$, Eq.\ (\ref{eq:BzD1D2}) is generically larger for the odd-parity OPs so that they can lower their free energy more efficiently in an applied magnetic field $B_z$.

We have argued above that in the absence of a magnetic field, $B_{1g+}$ pairing is favored, with a possibly coexisting $B_{1u+}$ superconducting OP (and thus AFM order). The $B_z$ field then induces $B_{2g+}$ and possibly $B_{2u+}$ pairing. The emerging scenario is thus that the larger free-{en\-er\-gy} gain due to the coexistence of the odd-parity $B_{1u+}$ and $B_{2u+}$ OPs stabilizes these states above a sufficiently strong magnetic field $B_z$, where the first-order transition takes place. It is then likely that the $B_{2u+}$ contribution is stronger than the $B_{1u+}$ contribution since $B_{2u+}$ is nearly degenerate with $B_{1g+}$. Note that in the presence of AFM order superconductivity involves the same four irreps $B_{1g+}$, $B_{2g+}$, $B_{1u+}$, and $B_{2u+}$ below and above the first-order transition. There is thus no change of symmetry. We will return to this point below.

Machida \cite{machida_violation_2022} proposed a magnetic mechanism for the superconductor-to-su\-per\-con\-duc\-tor transition in a field $B_z$: It is essentially a spin-flop transition of an antiferromagnet of local $\mathrm{Ce}$ moments in a magnetic field applied in parallel to the low-field N\'eel vector. In this scenario, the low-field superconducting phase has AFM order with N\'eel vector along the \textit{z}-axis, which is consistent with the $A_{1u-}$ magnetic order discussed above. At the first-order transition, the N\'eel vector jumps into the \textit{xy} plane and the $\mathrm{Ce}$ moments are canted by the field, leading to a nonzero magnetization \cite{machida_violation_2022}. Such a magnetically driven scenario could not explain a situation where the magnetic transition meets the first-order superconductor-to-superconductor phase transition below $T_c$ \cite{ogata_parity_2023}.

\subsection{In-plane field}
\label{subsec:coupling_in-plane_fields}

The in-plane field components $(B_x, B_y)$ transform according to $E_{g-}$. This requires a superconducting OP belonging to a two-dimensional irrep to construct a trilinear term. The two irreps $\Gamma_1$ and $\Gamma_2$ of the superconducting OPs must be such that the product $E_{g-} \otimes A_{1g-} \otimes \Gamma_1 \otimes \Gamma_2 = E_{g+} \otimes \Gamma_1 \otimes \Gamma_2$ contains $A_{1g+}$. Taking $\Gamma_1 = B_{1g+}$, $\Gamma_2$ has to be $E_{g+}$. The coexisting $B_{1u+}$ induces $E_{u+}$.

The allowed trilinear term has the same form for primary $B_{1g+}$ and $B_{1u+}$:
\begin{align}
F_3 &= i\, \big[ \Delta_1^*\, (B_x \Delta_{2x} - B_y \Delta_{2y})
  - \Delta_1\, (B_x \Delta_{2x}^* - B_y \Delta_{2y}^*) \big] ,
\label{eq:F3B1g.B1u.2}
\end{align}
where the two-dimensional $E_{g+}$ or $E_{u+}$ OP is written as $(\Delta_{2x},\Delta_{2y})$~\footnote{The convention for the components is the same as for the doublets $(\sigma_x,\sigma_y)$ and $(B_x,B_y)$ in the case of $E_g$ and the same multiplied by an $A_{1u}$ basis function in the case of $E_u$.}. Minimization of $F_3$ determines the orientation of the vector OP $(\Delta_{2x},\Delta_{2y})$ for given field $(B_x,B_y)$.

As discussed in Sec.\ \ref{sec:SC_pairing_states}, the $E_{g+}$ OP induced by $B_{1g+}$ is strongly suppressed at the Fermi energy so that we do not expect that the system gains significant free energy by inducing $E_{g+}$. The $E_{u+}$ OP induced by $B_{1u+}$ is not as strongly disfavored. However,  unlike the induced $B_{2u+}$ OP for out-of-plane field, it has no reason to be close in energy to the primary $B_{1g+}$ OP. This near-degeneracy is essential for the superconductor-to-superconductor phase transition. Since there is no near-degeneracy relevant for in-plane fields, we do not expect such a transition in this case and, in particular, no phase dominated by an $E_{u+}$ OP at strong fields.

\subsection{Pauli limiting}
\label{subsec:Pauli_limiting}

The most commonly discussed mechanism in the context of the superconductor-to-superconductor transition is Pauli (paramagnetic) limiting \cite{khim_field-induced_2021, cavanagh_nonsymmorphic_2022, machida_violation_2022, amin_kramers_2024, NaB24, ogata_parity_2023, szabo_effects_2024, schertenleib_unusual_2021}, i.e., the destruction of a superconducting state by an applied magnetic field due to the Zeeman splitting of the bands \cite{chandrasekhar_note_1962, clogston_upper_1962, WHH66}. We now discuss how Pauli limiting changes the picture developed in the previous sections.

For a $B_z$ field, the originally degenerate pseudospin bands, and thus the Fermi surface, in the normal state split into pseudospin-up and pseudospin-down bands. A pairing state is Pauli limited if it does not permit a weak-coupling instability by pairing electronic states at momenta $\mathbf{k}$ and $-\mathbf{k}$ on the split Fermi surface. We do not consider pairing of electrons the momenta of which are not opposite, i.e., Fulde--Ferrell--Larkin--Ovchinikov states \cite{fulde_superconductivity_1964, larkin_nonuniform_1965}, since there is no direct evidence for such states in CeRh\textsubscript{2}As\textsubscript{2}. However, an inhomogeneous state has recently been invoked as a way to explain a discrepancy between the Pauli limiting field obtained by extrapolating Knight shift data to zero temperature and the actual upper critical field, for the high-field phase~\cite{ogata_parity_2023}.

Pairing states are Pauli limited by a field $B_z$ if the Cooper pairs are made up of electrons with opposite pseudospin orientation in the standard (spin-\textit{z}) basis. On the one hand, this is the case for pseudospin-singlet states. On the other hand, pseudospin-triplet states are described by a pairing potential of the form $\mathbf{d}(\mathbf{k}) \cdot \mathbf{s}\, is_y$ \cite{sigrist_phenomenological_1991}. The \textit{z}-component of the $\mathbf{d}$ vector corresponds to $s_z\, is_y = s_x$ and thus describes the opposite-pseudospin contribution to triplet pairing, which is Pauli limited~\cite{chandrasekhar_note_1962, clogston_upper_1962}.

As noted in Sec.\ \ref{subsec:pseudospin}, even-parity (odd-parity) pairing strictly corresponds to pseudospin-singlet (pseudospin-triplet) pairing. Hence, the $B_{1g+}$ and $B_{2g+}$ pairing states are projected onto pseudospin-singlet states, whereas the $B_{1u+}$ and $B_{2u+}$ states are projected onto pseudospin-triplet states. Moreover, as shown in Table \ref{tab:SC_states}, the $\mathbf{d}$ vectors of the $B_{1u+}$ and $B_{2u+}$ states predominantly lie in the \textit{xy} plane~\cite{cavanagh_nonsymmorphic_2022}. Hence, the $B_{1g+}$ and $B_{2g+}$ states are Pauli limited. On the other hand, the $B_{1u+}$ and $B_{2u+}$ states, which we have argued to dominate in the high-field phase, mostly avoid Pauli limiting.

We thus find that the inclusion of Pauli limiting does not overturn our scenario but refines it. The main additional insight is that the odd-parity pairing states are not only favored over the even-parity ones by larger trilinear coupling terms but also by avoiding Pauli limiting.

At $B_z$ beyond about $7\, \mathrm{T}$, AFM order is absent \cite{khanenko_phase_2025}. The even-parity, pseudospin-singlet states are strongly suppressed by Pauli limiting. Of the low-energy partner states $B_{1g+}$ and $B_{2u+}$, only the latter avoids Pauli limiting. Hence, we expect strong $B_{2u+}$ pairing with a coexisting $B_{1u+}$ OP supported by the applied magnetic field. We briefly consider what would happen if superconducting states with $\pi$ phase shifts between layers ($\pi$ junctions) were favored, in contrast to what we have assumed in Sec.\ \ref{subsec:pairing}. Everything else being unchanged, the order of the nearly degenerate $B_{1g+}$ spin-singlet and $B_{2u+}$ spin-singlet states would be reversed, i.e., $B_{2u+}$ would become favored also at low fields $B_z$. In the pseudospin picture, this state corresponds to pseudospin-triplet pairing with $\mathbf{d}$ vector predominantly in plane, see Table \ref{tab:SC_states}, and would thus avoid Pauli limiting. Since $B_{2u+}$ pairing is also favored at high fields there would not be any reason for a first-order superconductor-to-superconductor transition.

For in-plane magnetic field, the $B_{1g+}$ state is again Pauli limited since it is a pseudospin-singlet state. The $B_{1u+}$ state induced by magnetic order is a pseudospin-triplet state with momentum-dependent $\mathbf{d}$ vector mostly in plane, see Table \ref{tab:SC_states}. The $\mathbf{d}$ vector is not orthogonal to the applied magnetic field for general momenta $\mathbf{k}$ and is thus subject to Pauli limiting~\cite{frigeri_superconductivity_2004}.

Moreover, $E_{g+}$ and $E_{u+}$ OPs are induced by an in-plane field. The first corresponds to pseudospin-singlet pairing and is thus Pauli limited, in addition to being strongly disfavored by the small renormalized gap. The $E_{u+}$ OP corresponds to pseudospin-triplet pairing with $\mathbf{d}$ vector strictly in the \textit{z}-direction, as Table \ref{tab:SC_states} shows.

In analogy to the previous discussion, Pauli limiting occurs for pseudospin-triplet states if the $\mathbf{d}$ vector contains a term parallel to the magnetic field $\mathbf{B}$. However here, the $\mathbf{d}$ vector is orthogonal to $\mathbf{B}$ so that the induced $E_{u+}$ OP avoids Pauli limiting. This is not expected to stabilize superconductivity at high in-plane fields for the reason discussed in the previous section: The $E_{u+}$ OP is not generically close in energy to the primary $B_{1g+}$ OP, unlike the $B_{2u+}$ OP.

\subsection{Implications of Knight shift data}
\label{sec:Knight_shift}

Recent NMR data \cite{ogata_parity_2023} showed clear signatures of the superconducting transitions in the Knight shift, which has implications for the superconducting states. For both the low-field and the high-field phase (at a single field value of $B_z = 4.5\,\mathrm{T}$, slightly above the first-order transition), the Knight shift shows a kink at $T_c$ and a decrease below $T_c$ with decreasing temperature. The magnetic transition in the low-field superconducting phase is invisible in the Knight shift for $\mathrm{As}(1)$, while the Knight shift for $\mathrm{As}(2)$ cannot be determined below this transition because of the very large line width. The absence of a signature for $\mathrm{As}(1)$ is consistent with an $A_{1u-}$ magnetic OP.

Generally, a temperature-dependent Knight shift below $T_c$ indicates a pseudospin-singlet state or a pseudospin-triplet state with $\mathbf{d}$ vector along the \textit{z}-di\-rec\-tion, for a $B_z$ field. Conversely, a pseudospin-triplet state with in-plane $\mathbf{d}$ vector is expected to lead to a temperature-independent Knight shift since there is no energy gap suppressing spin flips \cite{schrieffer_theory_1964}. Ogata \textit{et al.}\ \cite{ogata_parity_2023} conclude that both superconducting phases are what we would call pseudospin-singlet states. The possibility of a pseudospin-triplet state with $\mathbf{d}$ vector along the \textit{z}-direction is not considered. A primary $B_{1g+}$ superconducting OP in the low-field phase is consistent with the authors' interpretation. On the other hand, we have proposed coexisting $B_{1u+}$ and $B_{2u+}$ states in the high-field phase, which are inconsistent with the authors' interpretation.

It is unclear how any pseudospin-singlet state in the high-field regime could (mostly) avoid Pauli limiting. Moreover, the upper critical field is also inconsistent with the expectations based on the Knight-shift data. The Pauli-limiting field estimated from the $\mathrm{As}(1)$ Knight shift at $4.5\,\mathrm{T}$ is $4.8\,\mathrm{T}$ \cite{ogata_parity_2023}, much smaller than the actual upper critical field of about $14\,\mathrm{T}$. Ogata \textit{et al.}\ \cite{ogata_parity_2023} invoke a spatially modulated superconducting state in order to resolve the problem. In our scenario, the temperature-dependent Knight shift at $B_z = 4.5\,\mathrm{T}$ is naturally attributed to the coexistence of multiple superconducting OPs: there is still a sizable pseudospin-singlet contribution. In addition, the $\mathbf{d}$ vectors describing the pseudospin-triplet OPs do have a small \textit{z}-component. The singlet contribution dies out for increasing field, leaving only pseudospin-triplet pairing, which explains the high upper critical field.

\section{Landau free-energy expansion}
\label{sec:Landau_expansion}

In this section, we construct a phenomenological Landau expansion of the free energy. Based on this theory, we are able to generate phase diagrams that agree with state-of-the art experimental observations \cite{khim_field-induced_2021, ogata_parity_2023, landaeta_field-angle_2022, khanenko_origin_2025, kibune_observation_2022, khim_coexistence_2024, hafner_possible_2022, khanenko_phase_2025} and to analyze the complex interplay between magnetism, superconductivity, and an applied magnetic field in CeRh\textsubscript{2}As\textsubscript{2}.

We have seen that four superconducting OPs of $B_{1g+}$, $B_{2g+}$, $B_{1u+}$, and $B_{2u+}$ symmetry are nonzero in the presence of AFM order and an out-of-plane magnetic field. A Landau free energy involving all these OPs is a cumbersome expression. We therefore describe the superconductivity by two OPs, an even-parity one, which is large in the low-field phase, and an odd-parity one, which is large in the high-field phase. This can be achieved by starting from a Landau free energy in terms of all four superconducting OPs and integrating out two that are induced by the magnetic field in the presence of the other two, due to the trilinear invariants in Eqs.\ (\ref{eq:BzD1D2}) and (\ref{eq:F3B1g.B1u.2}). Specifically, the free energy is minimized first with respect to the two extra OPs and these OPs are then replaced everywhere by the solution in terms of the remaining OPs and the magnetic field. This procedure renormalizes the coefficients in the remaining free energy.

We choose $B_{1g+}$ and $B_{1u+}$ as the remaining, explicit OPs. $B_{1g+}$ is obvious since it is the preferred pairing state at zero field. The $B_{1u+}$ OP is then enforced by the AFM order, see Sec.\ \ref{subsec:coex_SC_M}. A trilinear term of the form of Eq.\ (\ref{eq:trilinear_coupling}) is thus allowed \footnote{If we were to write the free energy in terms of $B_{1g+}$ and $B_{2u+}$ OPs, then the trilinear term in Eq.\ (\ref{eq:trilinear_coupling}) would naively be forbidden by symmetry. At nonzero $B_z$, this trilinear term would be generated when the $B_{1u+}$ OP is integrated out. However, this mechanism does not work at zero field. The choice of $B_{1g+}$ and $B_{2u+}$ would thus fail to describe the trilinear coupling of even and odd superconducting OPs and the magnetic OP at zero field.}. Similar arguments can be made for the superconducting OPs induced by in-plane field components.

The expansion up to fourth order in symmetry-allowed products of $\Delta_1$ (irrep $B_{1g+}$), $\Delta_2$ ($B_{1u+}$), $M$ ($A_{1u-}$), $B_z$ ($A_{2g-}$), and $(B_x,B_y)$ ($E_{g-}$) can be written as
\begin{equation}
F = F_{\text{pure}} + F_{\text{mixed}}
  + F_{\text{field}} ,
\end{equation}
with
\begin{align}
F_{\text{pure}} &=
    \alpha_{1} \Delta_{1}^*\Delta_{1}
    + \alpha_{2} \Delta_{2}^*\Delta_{2}
    + \alpha_{M} M^2 \nonumber \\
&\quad{} + \beta_{1}\, (\Delta_{1}^*\Delta_{1})^2
    + \beta_{2}\, (\Delta_{2}^*\Delta_{2})^2
    + \beta_{M} M^4,
\label{eq:Landau_F_pure} \\
F_{\text{mixed}} &=
    \gamma_{12} \Delta_{1}^*\Delta_{1}\Delta_{2}^*\Delta_{2}
\nonumber \\
&\quad{} + \gamma_{1M} \Delta_{1}^*\Delta_{1} M^2
    + \gamma_{2M} \Delta_{2}^*\Delta_{2} M^2
\nonumber \\
&\quad{} + \delta_{12M} M i\, (\Delta_{1}^*\Delta_{2} - \Delta_{2}^*\Delta_{1}),
\label{eq:Landau_F_mixed} \\
F_{\text{field}} &= (B_{x}^2+B_{y}^2)\,
    (\lambda_{1}^{xy} \Delta_{1}^*\Delta_{1}
    + \lambda_{2}^{xy} \Delta_{2}^*\Delta_{2}
    + \lambda_{M}^{xy} M^2 ) \nonumber \\
&\quad{} + B_{z}^2\,
    (\lambda_{1}^{z} \Delta_{1}^*\Delta_{1}
    + \lambda_{2}^{z} \Delta_{2}^*\Delta_{2}
    + \lambda_{M}^{z} M^2 ) .
\label{eq:Landau_F_field}
\end{align}
Here, $F_{\text{pure}}$ contains standard terms of single OPs and the second-order coefficients are expanded to linear order in temperature,
\begin{align}
\alpha_1 &= \alpha_1'\, (T - T_{1}),
\label{eq:alpha1.T} \\
\alpha_2 &= \alpha_2'\, (T - T_{2}) ,
\label{eq:alpha2.T} \\
\alpha_M &= \alpha_M'\, (T - T_M) ,
\label{eq:alphaM.T}
\end{align}
with $\alpha_1', \alpha_2', \alpha_M > 0$, i.e., the second-order terms stabilize the corresponding OPs below their (bare) critical temperatures. Note that the actual critical temperature $T_c$ of superconductivity need not coincide with one of the temperatures $T_{1}$ and $T_{2}$ and that the actual N\'eel temperature $T_N$ need not coincide with $T_M$. Generally, we find $T_c \leq T_{1}$ at $B=0$ since in the case of $T_N > T_c$ magnetism slightly suppresses the superconducting order and shifts $T_c$ below $T_{1}$. For the same reason we have $T_N \leq T_M$, due to the suppression of magnetism by pre-existing superconductivity for $T_N < T_c$. In order for the free energy to be bounded, positive fourth-order terms are included.

In $F_{\text{mixed}}$, we introduce terms containing different OPs, i.e., couplings between ordered phases. The standard biquadratic terms carry coefficients $\gamma$. We want to highlight the distinctiveness of the trilinear coupling with coefficients $\delta_{12M}$, which guarantees that the third OP is automatically stabilized once the other two are present, see Eq.\ (\ref{eq:trilinear_coupling}) and the accompanying discussion above. This trilinear coupling will turn out to be of crucial importance since it enables, for example, the symmetry-preserving first-order superconductor-to-superconductor transition.

Furthermore, the coupling to an applied magnetic field is included in $F_{\text{field}}$, which involves only biquadratic terms. Recall that the effect of the field of inducing additional superconducting OPs has been integrated out.

To be specific, we assume $\mathbf{B} \parallel [110]$ for in-plane field orientation in accordance with \cite{ogata_appearance_2024, khanenko_origin_2025}, i.e., an in-plane angle of $\phi = \pi/4$, and parametrize the magnetic field as
\begin{equation}
\mathbf{B} = \begin{pmatrix}
    B_x \\ B_y \\ B_z
  \end{pmatrix}
  = B\, \begin{pmatrix} \sin\theta \cos\phi \\ \sin\theta \sin\phi \\ \cos\theta \end{pmatrix}
  = B\, \begin{pmatrix} \frac{\sin\theta}{\sqrt{2}}  \\ \frac{\sin\theta}{\sqrt{2}} \\ \cos\theta
  \end{pmatrix} .
\end{equation}
Since the relevant physical processes are expected to happen mostly deep inside the vortex phase, an extra term for field expulsion is disregarded. The coefficients $\lambda$ in $F_{\text{field}}$ are understood as effective quantities describing the leading-order limiting effects for the corresponding OPs by the magnetic field.

Regarding the general procedure, we obtain the coefficients of our model from a detailed analysis of the potential landscape and by fitting our model to experimental observations, while keeping in mind the insights gained in previous sections. The constraints on the coefficients will mostly be discussed in Sec.\ \ref{subsec:Landau_TN_above_TC}, which is concerned with the case $T_N > T_c$. The transition to the case $T_N < T_c$ in Sec.\ \ref{subsec:Landau_TN_below_TC} will merely involve a change of $T_M$.

\subsection{Onset of superconductivity inside the magnetic phase: $T_N > T_c$}
\label{subsec:Landau_TN_above_TC}

\begin{figure}
\centering
\includegraphics[width=.47\textwidth]{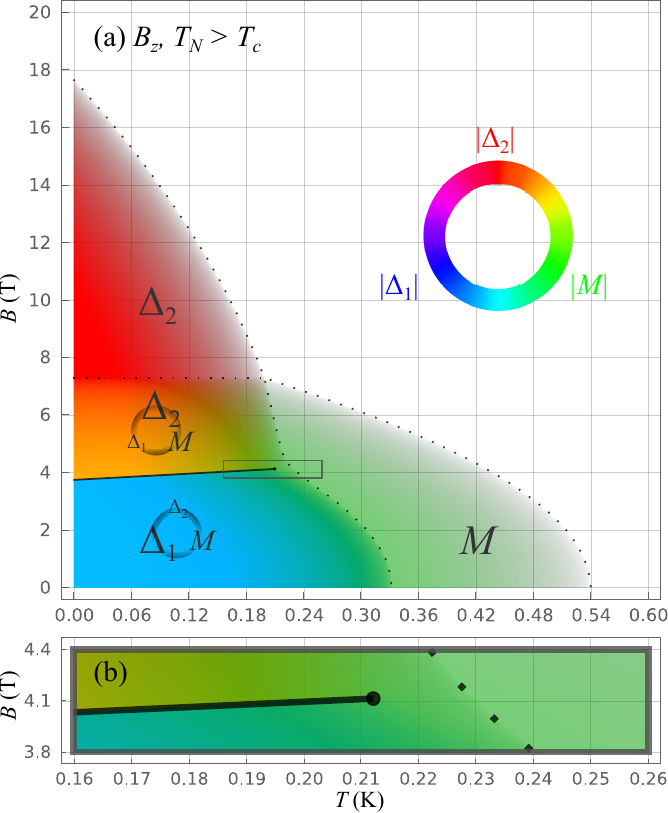}
\caption{Phase diagram for out-of-plane magnetic field and $T_N > T_c$ for (a) the full temperature and magnetic-field range and (b) the region close to the end point of the first-order transition. The colors of distinct phases are generated by mixing contributions from the even-parity OP $|\Delta_1|$ (blue), the odd-parity OP $|\Delta_2|$ (red), and the magnetic OP $|M|$ (green) into an RGB triplet. The transparency is determined from the overall OP magnitude. Solid and dotted lines represent first-order and second-order phase transitions, respectively.}
\label{fig:phase_diagram_and_inset_TN_above_out_of_plane}
\end{figure}

\begin{figure}
\centering
\includegraphics[width=.47\textwidth]{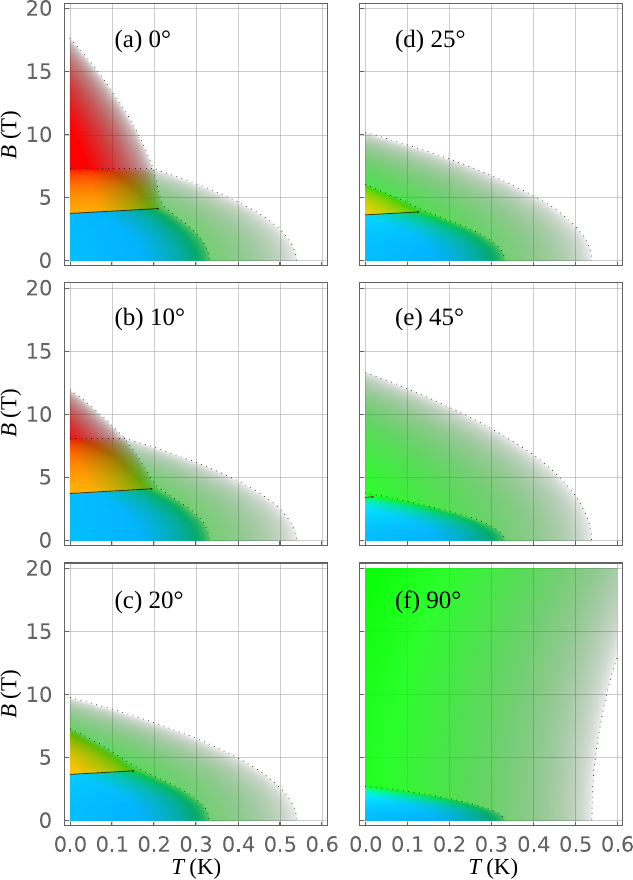}
\caption{Phase diagrams in the temperature-magnetic-field plane for $T_N > T_c$ and angles (a) $\theta = 0^{\circ}$, i.e., $\mathbf{B} \parallel [001]$, (b) $\theta = 10^\circ$, (c) $\theta = 20^\circ$, (d) $\theta = 25^\circ$, (e) $\theta = 45^\circ$, and (f) $\theta = 90^{\circ}$, i.e., $\mathbf{B} \parallel [110]$, between the magnetic field and the surface normal (\textit{z}-axis). The color reflects the admixture of distinct OPs as in Fig.~\ref{fig:phase_diagram_and_inset_TN_above_out_of_plane}. Solid and dotted lines represent first-order and second-order phase transitions, respectively.}
\label{fig:phase_diagram_TN_above_angular}
\end{figure}

As a first scenario, we investigate the case of $T_N > T_c$, i.e., we consider phase I introduced in Sec.\ \ref{sec:intro} to coincide with magnetic dipolar order, while superconductivity sets in at lower temperatures \cite{khim_coexistence_2024}. That is, we consider $T_N = T_0$. Experiments suggest a slightly lower transition temperature for $\Delta_2$ compared to $\Delta_1$ \cite{khim_field-induced_2021, landaeta_field-angle_2022, kibune_observation_2022, ogata_parity_2023}, hence we set $T_{1} = 0.34 \,\mathrm{K}$ and $T_{2} = 0.24 \,\mathrm{K}$. Moreover, we choose $T_{M} = 0.54 \,\mathrm{K} = T_N$. Note that due to the weak suppression of superconductivity by the magnetic, order we obtain $T_c \approx 0.33 \,\mathrm{K} \lesssim T_{1}$ at $B=0$.

The calculated phase diagram for out-of-plane field orientation is shown in Fig.~\ref{fig:phase_diagram_and_inset_TN_above_out_of_plane} and displays multiple phases, both superconducting and magnetic. There is a magnetic phase setting in via a second-order transition at $T_N$, which penetrates both into the high-field and into the low-field phase below the corresponding second-order superconducting transitions. Most intriguingly, our analysis not only shows a first-order transition at $B^* \approx 4\, \mathrm{T}$ between two coexistence phases involving nonzero $\Delta_1$, $\Delta_2$, and $M$ but also predicts that the first-order transition line has a critical end point. Hence, there is a small crossover region between the end point and the second-order superconducting transition line. Figure \ref{fig:phase_diagram_and_inset_TN_above_out_of_plane}(b) shows a magnification of the corresponding region. In the low-field coexistence phase, $\Delta_1$ and $M$ dominate, with only a small contribution from $\Delta_2$. In the high-field coexistence phase, the relative weights of $\Delta_1$ and $\Delta_2$ are reversed. At a stronger field of about $7\, \mathrm{T}$, the magnetic OP $M$ and the even-parity superconducting OP $\Delta_1$ vanish at a second-order transition and only the odd-parity superconducting OP $\Delta_2$ survives. The superconductivity breaks down only at much higher fields.

The trilinear coupling between superconducting and magnetic OPs with strength $\delta_{12M}$ in Eq.\ (\ref{eq:Landau_F_mixed}) is crucial here. In its absence, the phase diagram in Fig.\ \ref{fig:phase_diagram_and_inset_TN_above_out_of_plane} would be qualitatively different: Below the first-order transition, we would get even-parity $\Delta_1$ superconducting order coexisting with magnetic order, whereas above, we would get odd-parity $\Delta_2$ superconducting and magnetic order. Hence, below and above the first-order line, different symmetries would be broken. The line then could not have a critical end point and there would not be a crossover region. Instead, it would reach the superconducting $T_c$ line in a bicritical point.

In Fig.~\ref{fig:phase_diagram_TN_above_angular}, we show the angle-dependent evolution of the phase diagram from out-of-plane ($\mathbf{B} \parallel [001]$) to in-plane field ($\mathbf{B} \parallel [110]$) orientation. While for a small angle of $\theta = 10^\circ$ between the magnetic field and the \textit{z}-direction mostly the critical-field values change, already for $\theta = 20^\circ$, the pure, nonmagnetic $\Delta_2$ phase is completely suppressed. For $\theta = 45^\circ$, also the high-field coexistence phase \footnote{Recall that, unless we explicitly consider the case $\delta_{12M} = 0$, the term coexistence phase refers to finite amplitudes of all three OPs.} is almost entirely gone, eventually leaving only the low-field coexistence phase and a much enhanced magnetic phase for $\theta = 90^\circ$ (in plane). A strong enhancement of the critical temperature of phase I in an in-plane magnetic field is indeed observed experimentally \cite{hafner_possible_2022}. Note that we do not describe the transition between the phases I and II \cite{hafner_possible_2022, schmidt_anisotropic_2024, khanenko_origin_2025, mishra_anisotropic_2022}, see Sec.~\ref{sec:intro}.

We now take a closer look at the choice of coefficients for our Landau analysis. Without loss of generality, we set the positive fourth-order coefficients to $\beta_1 = \beta_2 = \beta_M = 1$. This amounts to a choice of units for the OPs. The second-order coefficients are expanded in temperature to linear order, see Eqs.\ (\ref{eq:alpha1.T})--(\ref{eq:alphaM.T}). For a magnetic field to be able to induce a transition between the superconducting states, i.e., between the even-parity OP $\Delta_{1}$ and the odd-parity OP $\Delta_2$, these states have to be close in energy. Therefore, we set $\alpha_2' \gtrsim \alpha_1'$, with a slight imbalance to account for the weak tilting of the almost horizontal first-order transition line. Moreover, since the magnetic phase is expected to have a comparatively weak effect we assume that $\alpha_{1,2}' > \alpha_M'$.

\begin{figure*}[t]
\centering
\includegraphics[width=.98\textwidth]{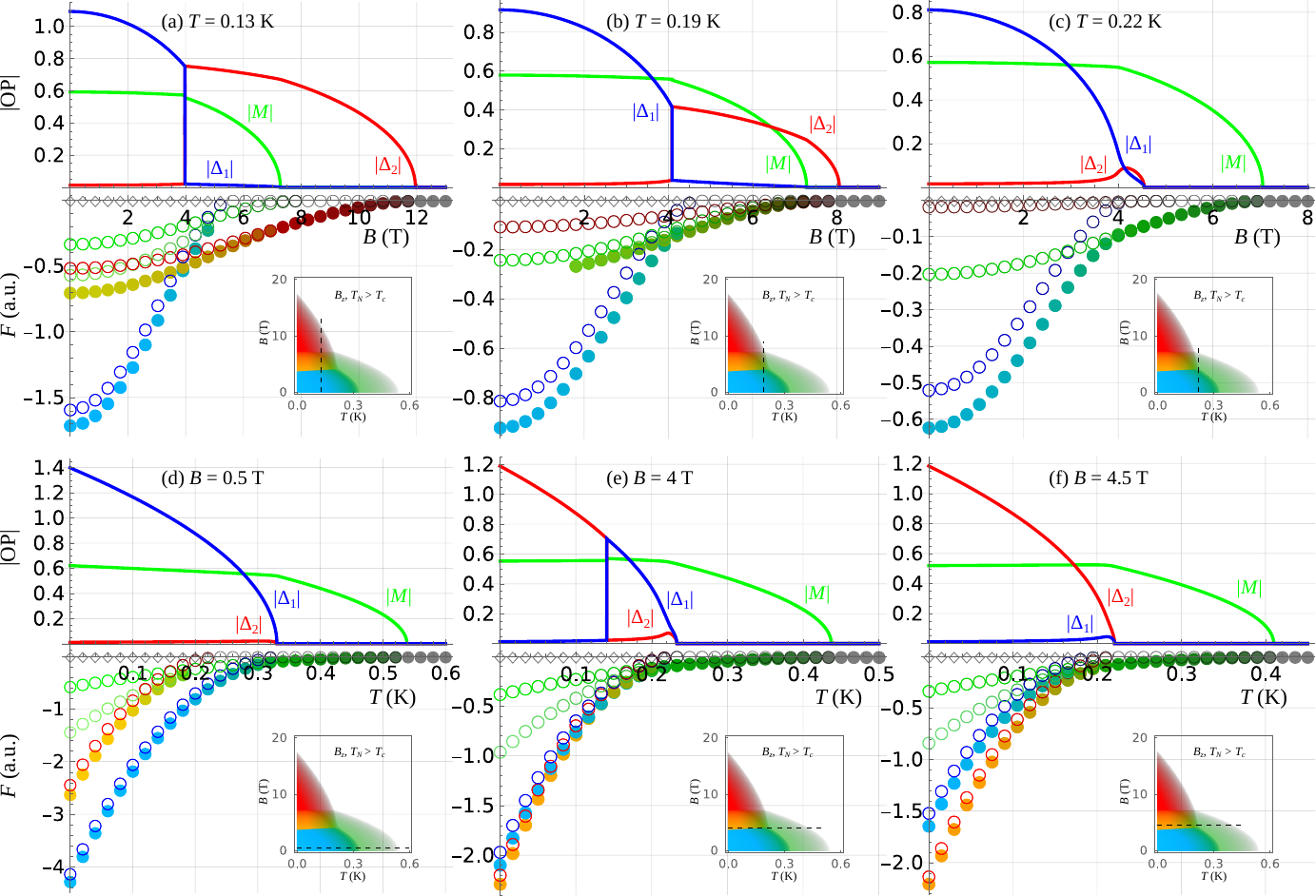}
\caption{Order-parameter amplitudes and free energies at stationary points along various cuts through the phase diagram for $T_N > T_c$ and magnetic field along the \textit{z}-direction. The cuts in the upper row are at a constant temperature of (a) $T = 0.13\,\mathrm{K}$, (b) $T = 0.19\,\mathrm{K}$, and (c) $T = 0.22\,\mathrm{K}$. The cuts in the lower row are at a constant magnetic field of (d) $B = 0.5\,\mathrm{T}$, (e) $B = 4\,\mathrm{T}$, and (f) $B = 4.5\,\mathrm{T}$. The cuts are also indicated in the insets. Minima, saddle points, and maxima of the free energy are denoted by the symbol $\bullet$, $\circ$, and $\diamond$, respectively. The color of the symbols reflects the admixture of distinct OPs, whose amplitudes are given in the corresponding upper part of the panels.}
\label{fig:combination_free_energy_OP_amplitudes_TN_above}
\end{figure*}

Next, we investigate the role of coupling terms between OPs. First, it is natural to assume a competition between distinct superconducting orders since they compete for the same electrons and thus $\gamma_{12}$ should be positive (repulsive). This helps to stabilize the first-order transition. Experimentally, the magnetic phase is slightly suppressed when entering the superconducting phases. For the low-field phase, this can be deduced from the suppression of internal fields $B_{\text{int}}$ below $T_c$ seen in $\mu$SR measurements \cite{khim_coexistence_2024}. For the high-field phase, the transition line of the magnetic transition deviates from a smooth extrapolation by being tilted toward lower fields. It becomes almost field independent inside the superconducting region \cite{khanenko_phase_2025}. This also indicates a weak suppression. It turns out that $\gamma_{12} \gg \gamma_{2M} \gtrsim \gamma_{1M} > 0$ gives best results. The trilinear coupling term is expected to be weak and we set $0<|\delta_{12M}|<1$. However, for the general phenomenology of a symmetry-preserving first-order superconductor-to-superconductor transition it only needs to be nonzero. Note that the sign of $\delta_{12M}$ is irrelevant, as the phase difference of $\pm \pi/2$ between the superconducting OPs will adjust such that Eq. (\ref{eq:trilinear_coupling}) reduces the free energy.

Finally, we discuss the biquadratic coupling terms involving the applied magnetic field. As discussed in Sec.\ \ref{subsec:Pauli_limiting}, the even-parity OP $\Delta_1$ experimentally shows reduced Pauli limiting for a magnetic field along the \textit{z}-direction, while the odd-parity OP $\Delta_2$ seems to avoid Pauli limiting completely. This scenario can be qualitatively modeled by choosing $1 \gg \lambda_{1}^z > \lambda_{M}^z \gtrsim \lambda_{2}^z > 0$.

Angle-resolved measurements of superconductivity in CeRh\textsubscript{2}As\textsubscript{2} have shown a strong variation of the upper critical field when tilting the field direction into the \textit{xy} plane \cite{landaeta_field-angle_2022}. Recall that we assume the in-plane field to be along the $[110]$ direction. In fact, the high-field phases completely vanish for in-plane fields, which we incorporate by choosing $\lambda_{2}^z \ll \lambda_{2}^{xy}$. This is not surprising since the odd-parity $B_{1u+}$ and $B_{2u+}$ OPs that dominate at large $B_z$ are pseudospin-triplet states with $\mathbf{d}$ vector predominantly in the \textit{xy} plane. This pairing is now suppressed by the in-plane field by Pauli limiting. Moreover, our analysis suggests a similar suppression of $\Delta_{1}$ and $\Delta_{2}$ for in-plane fields, hence $\lambda_{1}^{xy} \approx \lambda_{2}^{xy}$. That is, also the pseudospin-singlet $B_{1g+}$ and $B_{2g+}$ low-field pairing states face stronger suppression for in-plane fields. In Secs.\ \ref{subsec:coupling_out-of-plane_fields} and \ref{subsec:coupling_in-plane_fields}, we concluded that induced superconducting OPs can help to stabilize superconductivity for fields along the \textit{z}-direction but not for fields within the \textit{xy} plane~\footnote{We note that the inclusion of thus far disregarded potential in-plane ($E_{g-}$) FM magnetic contributions and their couplings to the superconducting OPs could yield an alternative explanation.}.

Apart from the positive coefficients mentioned above, we introduce a small but negative coefficient $\lambda_{M}^{xy} < 0$ for the coupling of the magnetic OP to an in-plane field since such a field was experimentally observed to stabilize phase I---recall that we identify this phase with magnetic order. In this context, the field-induced mixing of the $\Gamma_6$ excited doublet with the $\Gamma_7^{(1)}$ ground-state doublet has been discussed, see Fig.~\ref{fig:level_splitting}, leading to the stabilization of in-plane AFM order by field-induced dipole-quadrupole coupling \cite{schmidt_anisotropic_2024}. This results in a transition between multipolar phases I and II. This transition is not captured by our Landau functional, which does not contain higher multipolar OPs since our main focus lies on the superconducting states. Therefore, we effectively approximate the stabilization by an attractive coupling $\lambda_M^{xy} < 0$.

The complete set of coefficients is the following: ${\alpha_1' = 12}$, $\alpha_2' = 13$, $\alpha_M' = 2.8$, $\beta_1 = \beta_2 = \beta_M = 1$, $\gamma_{12} = 12$, $\gamma_{1M} = 0.38$, $\gamma_{2M} = 0.44$, $|\delta_{12M}| = 0.55$, $\lambda_1^z = 0.08$, $\lambda_2^z = 0.01$, $\lambda_M^z = 0.018$, $\lambda_1^{xy} = 0.5$, $\lambda_2^{xy} = 0.4$, and $\lambda_M^{xy} = -0.001$.

In Fig.~\ref{fig:combination_free_energy_OP_amplitudes_TN_above}, we show plots of the OP amplitudes and free energy for various cuts through the phase diagram, tuning either the magnetic field or the temperature. In panels (a) and (b), we observe a clear jump of $|\Delta_1|$ and $|\Delta_2|$ when crossing the first-order transition at $B^* \approx 4\,\mathrm{T}$, which is accompanied by a discontinuous change of slope of the lowest minimum of the free energy. However, all three OPs remain nonzero up to the second-order transition to a pure $\Delta_2$ phase at higher fields. At a slightly higher temperature of $T = 0.22\,\mathrm{K}$, panel (c), we observe that the first-order transition line is replaced by a smooth crossover. An experimental observation of such a crossover region would be a direct verification of our proposal of a symmetry-preserving first-order transition between two multicomponent superconducting phases. In panels (d)--(f), the aforementioned suppression of the magnetic OP when entering the superconducting region can be observed. Panel (e) once again shows a jump in OP amplitudes when crossing the first-order transition line, which is not completely horizontal.

\subsection{Onset of magnetism below the superconducting transition: $T_N < T_c$}
\label{subsec:Landau_TN_below_TC}

In this subsection, we discuss the scenario of $T_N < T_c$ based on NMR and NQR measurements \cite{ogata_parity_2023, kibune_observation_2022}. The main idea here is to assume that the coefficient $\alpha_M$ controlling the dynamical magnetic order depends on the experimental timescale \cite{khim_coexistence_2024}, as discussed in Sec.\ \ref{sec:magnetism}, whereas the superconducting properties do not. We effectively achieve this by changing a single parameter, namely the magnetic critical temperature $T_M$ from $T_M = 0.54 \,\mathrm{K}$ to $T_M = 0.32 \,\mathrm{K}$, which is slightly below $T_c = T_{1} = 0.34 \,\mathrm{K}$ but above $T_{2} = 0.24 \,\mathrm{K}$.

\begin{figure}[t!]
\centering
\includegraphics[width=.47\textwidth]{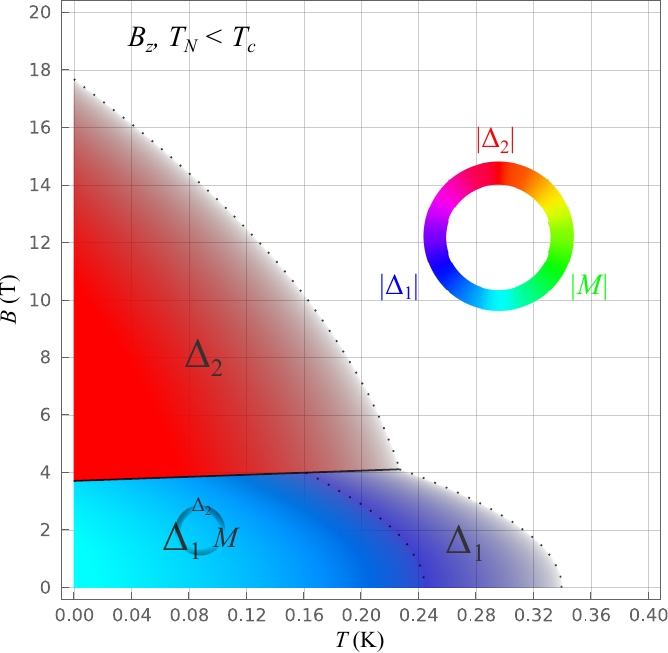}
\caption{Phase diagram in the temperature-magnetic-field plane for out-of-plane magnetic field and $T_N < T_c$. The colors of distinct phases are generated by mixing contributions from the even-parity OP $|\Delta_1|$ (blue), the odd-parity OP $|\Delta_2|$ (red), and the magnetic OP $|M|$ (green) into an RGB triplet. The transparency is determined from the overall OP magnitude. Solid and dotted lines represent first-order and second-order phase transitions, respectively.}
\label{fig:phase_diagram_TN_below_out_of_plane_dense}
\end{figure}

The overall phase diagram now changes dramatically. In particular, for out-of-plane fields, no stable purely magnetic phase is realized, as is shown in Fig.~\ref{fig:phase_diagram_TN_below_out_of_plane_dense}. The first transition that occurs for decreasing temperature at $B=0$ is a second-order superconducting transition at $T_c$ into a $\Delta_1$ phase, followed by another second-order transition at $T_N$, where TR symmetry breaks. Below $T_N$, the OPs $\Delta_1$, $\Delta_2$, and $M$ coexist. Note that at $B=0$ the actual N\'eel temperature $T_N$ resulting from our calculations lies closer to $T_{2}$ than to $T_M$, i.e, $T_M \neq T_N \gtrsim 0.24 \,\mathrm{K} = T_{2}$. This can be understood in terms of the suppression of magnetism by the superconducting phase $\Delta_1$, as well as the trilinear coupling between the three OPs which in turn facilitates a stabilization close to the onset of $\Delta_2$. Furthermore, the almost horizontal first-order transition now happens between either the pure $\Delta_1$-phase or the low-field coexistence-phase and a pure $\Delta_2$-phase at higher field. A high-field coexistence phase is not found for $B$ along the \textit{z}-direction. Note that the occurrence of an end point and crossover region is not possible in this case, since the first-order transition involves symmetry breaking.

\begin{figure}[t!]
\centering
\includegraphics[width=.47\textwidth]{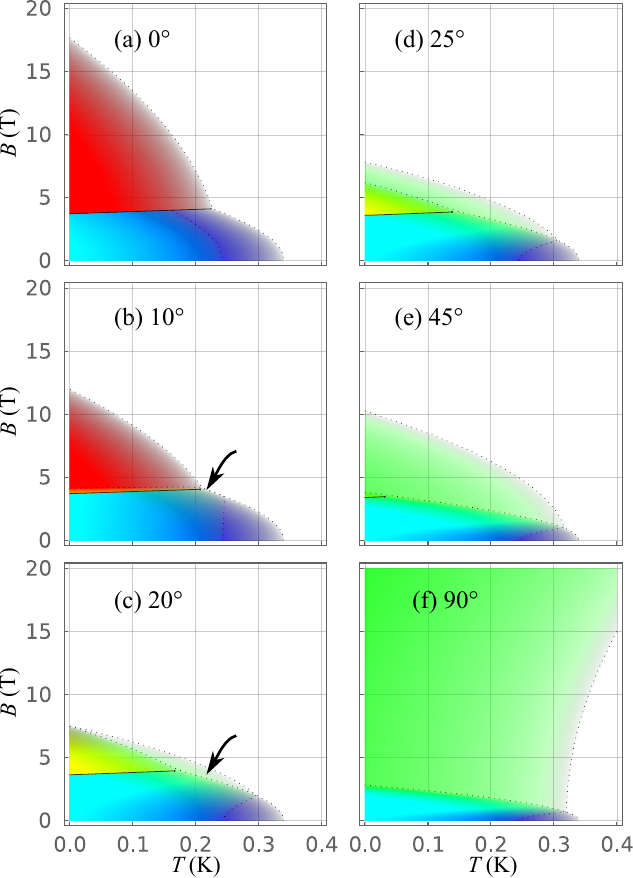}
\caption{Phase diagrams in the temperature-magnetic-field plane for $T_N < T_c$ and angles (a) $\theta = 0^{\circ}$, i.e., $\mathbf{B} \parallel [001]$, (b) $\theta = 10^\circ$, (c) $\theta = 20^\circ$, (d) $\theta = 25^\circ$, (e) $\theta = 45^\circ$, and (f) $\theta = 90^{\circ}$, i.e., $\mathbf{B} \parallel [110]$, between the magnetic field and the surface normal (\textit{z}-axis). The color reflects the admixture of distinct OPs as in Fig.~\ref{fig:phase_diagram_TN_below_out_of_plane_dense}. Solid and dotted lines represent first-order and second-order phase transitions, respectively.}
\label{fig:phase_diagram_TN_below_angular}
\end{figure}

\begin{figure}
\centering
\includegraphics[width=.47\textwidth]{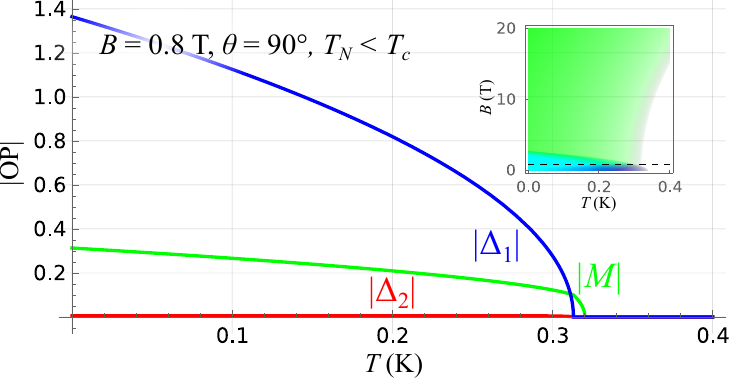}
\caption{Order-parameter amplitudes for in-plane field ($\mathbf{B} \parallel [110]$) and $T_N < T_c$ at constant field $B = 0.8 \,\mathrm{T}$. Here, $\Delta_1$ and $M$, and therefore also $\Delta_2$, order at closely proximate temperatures.}
\label{fig:OP_amplitudes_TN_below_angle_90_B0Point5}
\end{figure}

If the applied field is tilted away from the \textit{z}-direction, the high-field phase vanishes between angles $\theta=10^\circ$ and $\theta=20^\circ$. Moreover, a narrow region of purely magnetic OP sets in at $\theta=10^\circ$ and is clearly visible at $\theta=20^\circ$. This region is marked by arrows in Figs.~\ref{fig:phase_diagram_TN_below_angular}(b) and (c), respectively. Therefore, the absence of symmetry breaking again enables the appearance of a small crossover region between two coexistence phases and an end point of the first-order transition line. Analogous to the case $T_N > T_c$, setting $\delta_{12M} = 0$ suppresses the induction of the third OP by the other two and thus enforces a symmetry-breaking transition. Furthermore, as in the case $T_N > T_c$, the pure $\Delta_2$ state is absent at an angle of $20^\circ$. This evolution continues for larger angles $\theta$, until for $\theta = 90^\circ$ the magnetic OP is stable over a broad region of the phase diagram. However, in contrast to $T_N > T_c$, we still find a finite region of pure $\Delta_1$ superconducting order since at $B=0$, $T_N < T_c$ remains true for all angles. We note an interesting observation for $\theta=90^\circ$: the magnetic and superconducting transitions almost exactly coincide at $B = 0.8 \,\mathrm{T}$, the field strength at which recent NMR measurements with $\textbf{B} \parallel [110]$ were performed \cite{ogata_appearance_2024}. This can be seen in Fig.~\ref{fig:OP_amplitudes_TN_below_angle_90_B0Point5} and could be an alternative explanation for the nearly simultaneous onset of both line-width broadening and superconductivity, as opposed to the case of magnetic field along the \textit{z}-direction~\cite{ogata_parity_2023, kibune_observation_2022}.

It is quite remarkable that both topologies of the phase diagram for out-of-plane magnetic field, i.e., $T_N>T_c$ and $T_N<T_c$, can be obtained by tuning a single parameter. We find that respecting a broad range of known experimental features for both scenarios and for variable field angle imposes strong constraints on the choice of parameters. For example, a weaker repulsion $\gamma_{12}$ between the even-parity and odd-parity superconducting OPs would lead to the penetration of magnetic order into the high-field phase also for $T_N<T_c$ and thus a high-field coexistence phase for magnetic field along the \textit{z}-direction, which is not suggested experimentally \cite{ogata_parity_2023}. Given the present parameter set, the constraint on $\gamma_{12}$ is less relevant for $T_N>T_c$, where also smaller $\gamma_{12}$ would suffice. However, a scenario of a very narrow high-field coexistence region for $T_N<T_c$ would also be in agreement with available NMR data \cite{ogata_parity_2023}.

\begin{figure*}
\centering
\includegraphics[width=.98\textwidth]{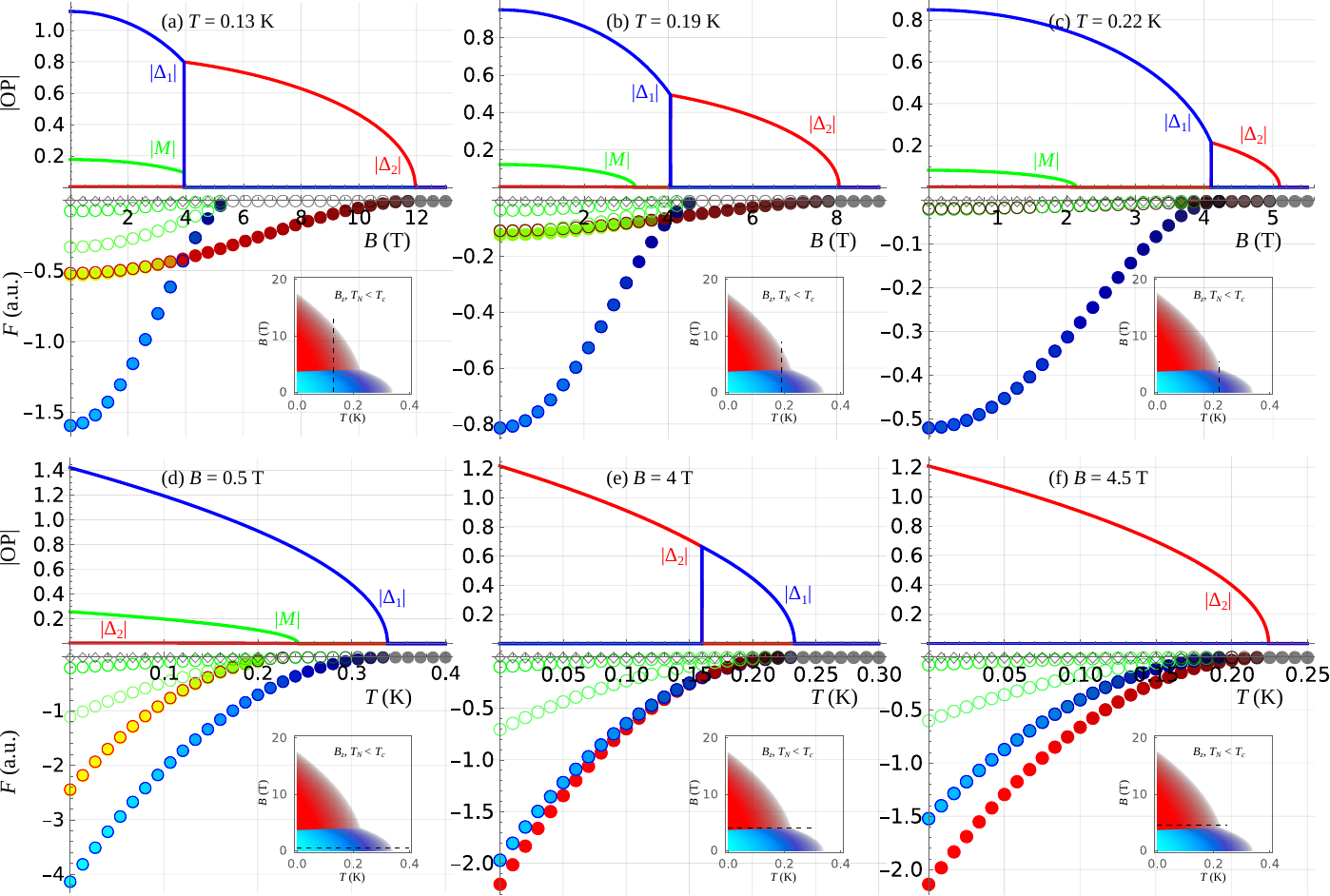}
\caption{Order-parameter amplitudes and free energies at stationary points along various cuts through the phase diagram for $T_N < T_c$ and magnetic field along the \textit{z}-direction. The cuts in the upper row are at a constant temperature of (a) $T = 0.13\,\mathrm{K}$, (b) $T = 0.19\,\mathrm{K}$, and (c) $T = 0.22\,\mathrm{K}$. The cuts in the lower row are at a constant magnetic field of (d) $B = 0.5\,\mathrm{T}$, (e) $B = 4\,\mathrm{T}$, and (f) $B = 4.5\,\mathrm{T}$. The cuts are also indicated in the insets. Minima, saddle points, and maxima of the free energy are denoted by the symbol $\bullet$, $\circ$, and $\diamond$, respectively. The color of the symbols reflects the admixture of distinct OPs, whose amplitudes are given in the corresponding upper part of the panels.}
\label{fig:combination_free_energy_OP_amplitudes_TN_below}
\end{figure*}

Figure \ref{fig:combination_free_energy_OP_amplitudes_TN_below} shows the evolution of the OP amplitudes and free energy for various cuts through the phase diagram. In panel (a) for constant $T = 0.13\,\mathrm{K}$, a large jump of all three OPs at the first-order transition can be seen. The high-field phase here only contains a nonzero OP $\Delta_2$. Panels (b) and (c) for higher temperatures instead show a disappearance of the magnetic OP and therefore also of $\Delta_2$ before reaching the first-order line. Hence, this first-order transition separates pure $\Delta_1$ and $\Delta_2$ phases, i.e., phases of different symmetry. Therefore, the first-order transition line cannot end in a critical point, unlike for $T_N > T_c$.

Due to the stabilization of magnetic order at lower temperatures and the competition with the pre-existing OP $\Delta_1$, the magnetic order realizes much smaller amplitudes compared to $T_N > T_c$. Therefore, $\Delta_2$ is extremely small in the coexistence phase since it also competes with $\Delta_1$ and the stabilization due to the trilinear coupling (\ref{eq:trilinear_coupling}) is weak. At the first-order transition ($B^* \approx 4\,\mathrm{T}$), the first derivative of the free energy again jumps when one minimum becomes disfavored with respect to another already existing minimum. For $T_N < T_c$, most of the minima are accompanied by almost degenerate saddle points, denoted by open circles in Fig.\ \ref{fig:combination_free_energy_OP_amplitudes_TN_below}. Panel (d) shows the second-order transition into the $\Delta_1$ phase and subsequently into the low-field coexistence phase. Like in the case of $T_N > T_c$, a discontinuity of $|\Delta_1|$ and $|\Delta_2|$ occurs when the cut intersects the slightly tilted first-order line in panel (e). Panel (f) only shows the mean-field onset of the high-field order $\Delta_2$.

\section{Thermodynamics}
\label{sec:thermodynamics}

\begin{figure*}
\centering
\includegraphics[width=.98\textwidth]{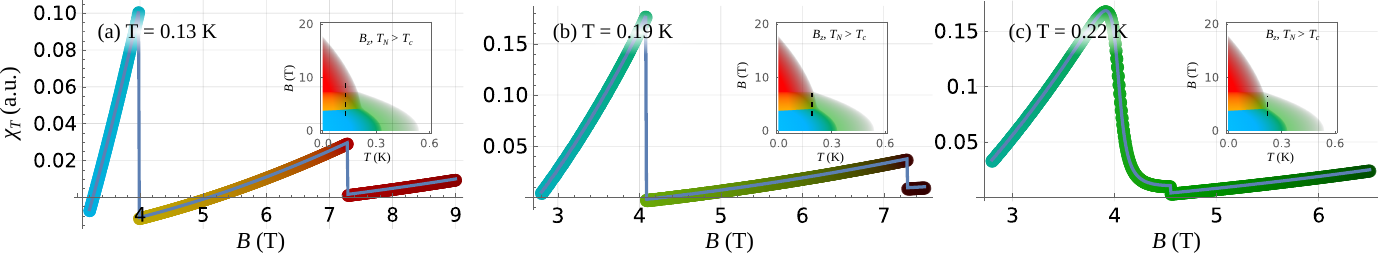}
\caption{Isothermal susceptibility $\chi_T$ for $T_N > T_c$ and $\mathbf{B} \parallel [001]$ as a function of magnetic field at (a) $T = 0.13\,\mathrm{K}$, (b) $T = 0.19\,\mathrm{K}$, and (c) $T = 0.22\,\mathrm{K}$. The corresponding cuts through the phase diagram are indicated in the insets. In panels (a) and (b), $\chi_T$ shows a jump at the first-order transition near $B^* \approx 4\, \mathrm{T}$, as well as at the second-order transition near $B_0 \approx 7\, \mathrm{T}$. Panel (c) shows a smooth transition in the crossover region near $B^*$ and a kink at $B_{c2} \approx 4.5\, \mathrm{T}$.}
\label{fig:susceptibility_TN_above}
\end{figure*}

\begin{figure*}
\centering
\includegraphics[width=.98\textwidth]{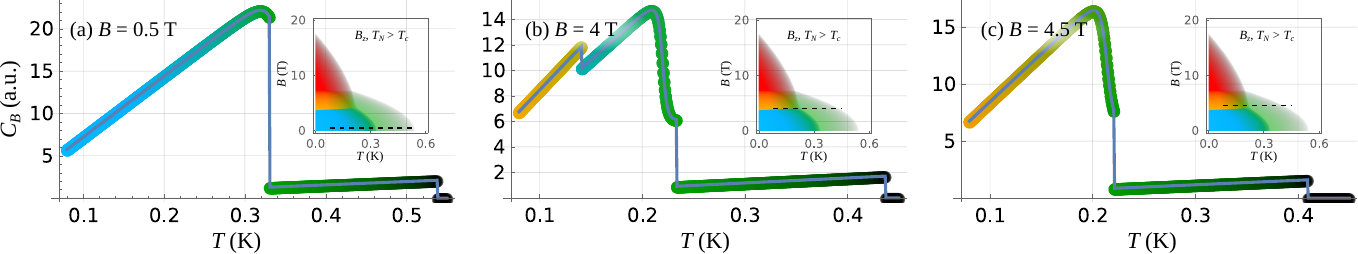}
\caption{Specific heat $C_B$ as a function of temperature for $T_N > T_c$ and magnetic field (a) $B = 0.5\,\mathrm{T}$, (b) $B = 4\,\mathrm{T}$, and (c) $B = 4.5\,\mathrm{T}$ along the \textit{z}-direction. The cuts are indicated in the insets. In all panels, a small jump is visible at $T_N$, which at lower temperature is followed by a large jump at $T_c$. Both transitions are of second order. Panel (b) additionally shows a small kink at $T^*$ when intersecting the first-order superconductor-to-superconductor transition.}
\label{fig:specific_heat_TN_above}
\end{figure*}

With the Landau free energy at hand, we can estimate experimentally relevant derivatives, e.g., thermodynamic quantities such as the specific heat or the susceptibility. In this section, we will focus on the case of $T_N > T_c$ and a magnetic field oriented along the \textit{z}-direction. Note that Landau theory, which involves an expansion in small OPs, is not expected to give quantitatively accurate results for temperatures far below $T_c$ and $T_N$.

Furthermore, our Landau functional does not contain the energy cost of the flux expulsion and thus does not account for the transition from the Mei\ss{}ner to the Shubnikov phase. Note that the corresponding lower critical field is much smaller than the characteristic field scale of the first-order superconductor-to-superconductor transition. Therefore, when calculating derivatives with respect to the magnetic field, we miss the contribution from flux expulsion. Nevertheless, we find qualitative agreement with recent experimental data for the uniform, isothermal susceptibility
\begin{align}
    \chi_T = -\left( \frac{\partial^2 F}{\partial B^2} \right)_{\!T} .
\end{align}
In Fig.~\ref{fig:susceptibility_TN_above}, $\chi_T$ is plotted as a function of magnetic field for three temperatures. For $T=0.13\,\mathrm{K}$ and $T=0.19\,\mathrm{K}$, the path in the phase diagram intersects the first-order transition at $B^* \approx 4\, \mathrm{T}$, indicated by a clear jump, followed by a smaller jump when the pure high-field superconducting state is reached at around $B_0 \approx 7\, \mathrm{T}$. Moreover, we observe a steeper slope of $\chi_T$ vs.\ $B$ below $B^*$ than above. These observations are compatible with, for instance, the data shown in {Fig.~S5} of Ref.\ \cite{khanenko_phase_2025}. At $T = 0.22\,\mathrm{K}$, we see that near $B^*$ the susceptibility smoothly changes with the field and no phase transition is observable, which is a hallmark of the underlying crossover region. Such behavior might be difficult to distinguish from an experimentally broadened phase transition and could be hidden in previously reported data. The smooth behavior of $\chi_T$ continues until the upper critical field is reached, indicated by a small kink near $B_{c2} \approx 4.5\, \mathrm{T}$.

The specific heat at constant magnetic field is given by
\begin{align}
    C_B = -T \left( \frac{\partial^2 F }{\partial T^2} \right)_{\!B} .
\end{align}
In Fig.~\ref{fig:specific_heat_TN_above}, we plot $C_B$ as a function of temperature for three values of the magnetic field. In all cases, a small jump is visible at the magnetic transition at $T_N$ and a larger jump at $T_c$. Both transitions are of second order. Panel (b) for $B = 4\,\mathrm{T}$ additionally shows a small kink at $T^* \approx 0.14\, \mathrm{K}$, the intersection with the first-order superconductor-to-superconductor transition, similar to {Fig.~3(a)} of Ref.\ \cite{khanenko_phase_2025}. Note that the cut in panel (b) but also the one in panel (c) passes quite close to the predicted critical end point at about $0.21\,\mathrm{K}$, see also Fig.~\ref{fig:phase_diagram_and_inset_TN_above_out_of_plane}. We interpret the pronounced maximum in $C_B$ as a fingerprint of the critical end point. In addition, we observe a small upturn in $C_B$ immediately below $T_c$ in panel (b). Its extent is primarily determined by and increases with the trilinear coupling coefficient $\delta_{12M}$. Intriguingly, this kink-like anomaly has actually been observed experimentally, see {Figs.\ 1 and 2} of Ref.\ \cite{chajewski_discovery_2024}. The authors interpret it as an additional first-order phase transition within the superconducting state. However, the anomaly was not observed in the specific heat below $B \approx 3\,\mathrm{T}$ and above $B \approx 5\,\mathrm{T}$ \cite{chajewski_discovery_2024}. This behavior agrees with our prediction that the anomaly should only appear in the vicinity of the critical end point.

Our calculations further indicate that the experimental signatures when crossing the superconductor-to-superconductor transition are indeed consistent with a first-order transition. This helps to resolve the recently raised concern regarding the order of this transition~\cite{khanenko_phase_2025}.

\section{Summary and conclusions}
\label{sec:summary}

Materials lacking inversion symmetry, i.e., noncentrosymmetric materials, often display unique magnetic and magnetoelectric behavior and potentially support topological superconductivity \cite{Schnyder_2015}. Lately, significant interest has emerged in crystal structures that break inversion symmetry locally but not globally. Superconductivity in such locally noncentrosymmetric materials can exhibit phenomena that would not usually be expected in centrosymmetric systems~\cite{FLS11, MSY12, SAF14, fischer_superconductivity_2023}.

One important class of locally noncentrosymmetric crystal structures consists of layered materials characterized by inversion centers between the layers but not within them. The tetragonal heavy-fermion superconductor CeRh\textsubscript{2}As\textsubscript{2} is a nonsymmorphic representative of this class, in which two layers per unit cell form two $\mathrm{Ce}$ sublattices. Another example is the family of superconductors based on the also nonsymmorphic structure BiS$_{2-n}$Se$_n$, including LaO$_{0.5}$Fe$_{0.5}$BiSe$_2$ \cite{fischer_superconductivity_2023, hoshi_extremely_2022}. Both CeRh\textsubscript{2}As\textsubscript{2} and $\mathrm{LaO_{0.5}Fe_{0.5}BiSe_2}$ have point group $D_{4h}$ and space group $P4/nmm$. The quasitetragonal cuprate $\mathrm{Bi_2Sr_2CaCu_2O_{8+\delta}}$ \cite{fischer_superconductivity_2023, miles_refinement_1998, gotlieb_revealing_2018, atkinson_microscopic_2020, lu_spin_2021} is a symmorphic example with point group $D_{2h}$ and space group $Bbmb$.

In this work, we have addressed the open question of probable symmetries of the ordered phases in the rich temperature-magnetic-field phase diagram of CeRh\textsubscript{2}As\textsubscript{2}, taking into account experimental data from NQR \cite{kitagawa_two-dimensional_2022}, NMR \cite{ogata_appearance_2024, ogata_parity_2023}, $\mu$SR \cite{khim_coexistence_2024}, and thermodynamic measurements \cite{landaeta_field-angle_2022, hafner_possible_2022, khim_field-induced_2021, khanenko_origin_2025, khanenko_phase_2025}. Within a theoretical framework based on symmetry analysis applied to a Bogoliubov--de Gennes Hamiltonian and Landau methods, we have made concrete predictions regarding the symmetry of superconducting and magnetic ordering phenomena, as well as their interaction.

As proposed earlier \cite{khim_field-induced_2021, landaeta_field-angle_2022, cavanagh_nonsymmorphic_2022, ogata_parity_2023, NaB24}, the superconducting pairing is predominantly of even parity at low fields but of odd parity for strong magnetic field applied along the \textit{z}-direction. In agreement with Refs.\ \cite{nogaki_even-odd_2022, amin_kramers_2024, lee_unified_2024, NaB24}, we argue that superconducting OPs belonging to $B$ irreps are preferred over OPs belonging to $A$ irreps because $A_{1g}$ and $A_{2u}$ orders are suppressed by the expected strongly repulsive Hubbard repulsion $U$ in the spin-singlet channel for this heavy-fermion compound.

An important perspective resulting from our calculations is that of not only multiphase but also multicomponent superconductivity in CeRh\textsubscript{2}As\textsubscript{2}. Here, we observe the generic coexistence of the even-parity OPs of $B_{1g+}$ and $B_{2g+}$ symmetry, which dominate at low fields, and the odd-parity $B_{1u+}$ and $B_{2u+}$ OPs, which dominate at high fields, in multiple regions of the phase diagram. This coexistence is enabled by trilinear invariants coupling the superconducting OPs and the AFM OP of $A_{1u}$ symmetry as well as between superconducting OPs and the out-of-plane magnetic field. We argue within a low-energy pseudospin picture that the invariant involving the \textit{z}-component of the applied magnetic field favors the odd-parity $B_{1u+}$ and $B_{2u+}$ OPs over the even-parity ones, eventually leading to the characteristic first-order superconductor-to-superconductor transition. Moreover, our scenario predicts this first-order transition to preserve symmetry---in the low-field and high-field coexistence phases, the same superconducting and magnetic OPs are realized, albeit with different amplitudes. One manifestation of this phenomenology is the observed critical end point of the first-order line when approaching the second-order superconducting transition, followed by a crossover region. Another important part of the mechanism is that an applied magnetic field suppresses even-parity, pseudospin-singlet pairing states due to Pauli limiting.

Our approach consistently reproduces experimental results for both $T_N > T_c$ and $T_N < T_c$, i.e., for both orderings of the zero-field critical temperatures of magnetic order and superconductivity. Phase diagrams of both types have been reported based on different experimental probes \cite{kibune_observation_2022, ogata_appearance_2024, khim_coexistence_2024}, perhaps due to the interplay of dynamical magnetic order and experimental timescales \cite{khim_coexistence_2024}. Notably, it is sufficient to assume a single parameter, namely, the critical temperature $T_M$, to depend on the experimental probe. Our approach also correctly describes the phase diagrams for general angles $\theta$ between the magnetic field and the surface normal.

One might question whether we are ``fitting an elephant'' since the Landau functional contains a large number of parameters. This is not so. One parameter per OP can be chosen arbitrarily up to the sign by simply rescaling the OP; we choose to set the $\beta$ parameters of fourth-order terms to unity. Reproducing the three-dimensional phase diagrams in the space of temperature, out-of-plane field, and in-plane field for both $T_N > T_c$ and $T_N < T_c$ strongly constrains the remaining parameters. The fact that this is possible at all thus suggests that it is based on a valid physical picture.

Our work answers open questions regarding the order of the superconductor-to-superconductor phase transition, as well as questions regarding the symmetry of competing orders, including a refined picture of their interplay. In particular, our proposal of multicomponent superconductivity and its implications redirect research on CeRh\textsubscript{2}As\textsubscript{2} away from the paradigm of single-OP phases. Moreover, a symmetry preserving first-order superconductor-to-superconductor transition represents a novel discovery in the context of unconventional superconductivity and CeRh\textsubscript{2}As\textsubscript{2} is a promising candidate to exhibit this phenomenon. Our calculations of thermodynamic observables suggest that signatures of such a transition---involving a critical end point and a crossover regime---may in fact be identified in available experimental data. At a more general level, our findings show the importance of symmetry-allowed but nontrivial couplings between superconducting and magnetic OPs, represented by trilinear terms in the Landau functional. In the present case, they lead to a strong connection between local inversion-symmetry breaking, even-parity and odd-parity superconductivity, as well as magnetic order.

\vspace*{3ex}

\begin{acknowledgments}
The authors wish to thank M. Brando, P. M. R. Brydon, C. Geibel, E. Hassinger, P. Khanenko, S. Khim, J. F. Landaeta, B. K. Nally, M. Pfeiffer, A. Ramires, B. Schmidt, A. L. Szab\'{o}, and K. Semeniuk for useful discussions and sharing of data. We are grateful to P. M. R. Brydon for pointing out a mistake in an earlier version of this paper. Financial support by Deutsche Forschungsgemeinschaft, in part through Collaborative Research Center SFB 1143, project A04, project id 247310070, and W\"urzburg--Dresden Cluster of Excellence ctd.qmat, EXC 2147, project id 390858490, is gratefully acknowledged.
\end{acknowledgments}

\section*{Data Availability}

The data that support the findings of this article are publicly available \cite{jakubczyk_data_2026}.

\appendix

\section{Pseudospin formalism}
\label{app:pseudospin}

In this appendix, the transformation of the BdG Hamiltonian in Eq.\ (\ref{1.HBdG.2}) into the pseudospin basis is reviewed. This representation is useful for the analysis of low-energy properties. In the absence of perturbations by a magnetic field or magnetic order, the transformed BdG Hamiltonian reads as
\begin{equation}
\tilde{\mathcal{H}}(\mathbf{k}) = \begin{pmatrix}
    \mathcal{H}_{++}(\mathbf{k}) & \mathcal{H}_{+-}(\mathbf{k}) \\
    \mathcal{H}_{+-}^\dagger(\mathbf{k}) & \mathcal{H}_{--}(\mathbf{k})
  \end{pmatrix} ,
\end{equation}
with the blocks
\begin{align}
\mathcal{H}_{\pm\pm}(\mathbf{k}) &= \begin{pmatrix}
    \xi_\pm(\mathbf{k})\, s_0 & \Delta_{\pm\pm}(\mathbf{k}) \\
    \Delta_{\pm\pm}^\dagger(\mathbf{k}) & -\xi_\pm(\mathbf{k})\, s_0
  \end{pmatrix} , \\*[0.5ex]
\mathcal{H}_{+-}(\mathbf{k}) &= \begin{pmatrix}
    0 & \Delta_{+-}(\mathbf{k}) \\
    \Delta_{-+}^\dagger(\mathbf{k}) & 0
  \end{pmatrix} ,
\end{align}
where $s_0$ is the identity matrix on pseudospin space. The normal-state dispersion $\xi_\pm(\mathbf{k}) = c_0(\mathbf{k}) \pm |\vec c(\mathbf{k})|$, where $\vec c = (c_1, c_2, c_3, c_4, c_5)$, is even in $\mathbf{k}$. The bands in the normal state are still twofold degenerate and can therefore be labeled by a pseudospin of length $1/2$ \cite{Fu15, ABT17, brydon_bogoliubov_2018, cavanagh_nonsymmorphic_2022}. For a natural choice of the pseudospin basis, the manifestly covariant Bloch basis \cite{Fu15, cavanagh_nonsymmorphic_2022}, the pseudospin transforms under symmetry operations like the real electron spin. For our model, the unitary matrix $\mathcal{U}$ implementing the transformation $\tilde{\mathcal{H}} = \mathcal{U}\mathcal{H}\mathcal{U}^\dagger$ can be expressed in terms of the coefficients $c_n(\mathbf{k})$ but the result is not particularly illuminating.

Without loss of generality, we consider the case that the band with dispersion $\xi_-(\mathbf{k})$ crosses the Fermi energy. At order zero in $\mathcal{H}_{+-}$, i.e., in the interband pairing, the system is just described by the $4\times 4$ BdG Hamiltonian $\mathcal{H}_{--}$. Its superconducting block, which describes \emph{intraband} pairing, can be written as $\Delta_{--}(\mathbf{k}) = \psi(\mathbf{k})\, is_y$ for pseudospin-singlet pairing and as $\Delta_{--}(\mathbf{k}) = \mathbf{d}(\mathbf{k})\cdot \mathbf{s}\, is_y$ for pseudospin-triplet pairing \cite{sigrist_phenomenological_1991}, where $\mathbf{s} = (s_x, s_y, s_z)$ are the Pauli matrices on pseudospin space. The matrix $\Delta_{--}(\mathbf{k})$ or, equivalently, the functions $\psi(\mathbf{k})$ and $\mathbf{d}(\mathbf{k})$, can be obtained in closed form using $\mathcal{U}(\mathbf{k})$, the expressions are rather lengthy. It is more illuminating to consider limiting cases, which is done in Sec.~\ref{subsec:pseudospin}.

It is important to realize that for the pseudospin-$1/2$ model, like for a real spin $1/2$, even-parity (odd-pa\-ri\-ty) pairing strictly corresponds to pseudospin-singlet (pseudospin-triplet) pairing. Hence, all superconducting OPs transforming according to ``\textit{g}'' (``\textit{u}'') irreps are mapped onto pseudospin-singlet (pseudospin-triplet) states.

The splitting of the quasiparticle bands is controlled by the TR-odd part $i(\mathbf{d} \times \mathbf{d}^*) \cdot \mathbf{s}$ of the gap product, which for pseudospin-triplet pairing reads as~\cite{sigrist_phenomenological_1991}
\begin{equation}
\Delta_{--}(\mathbf{k})\, \Delta_{--}^\dagger(\mathbf{k}) = |\mathbf{d}(\mathbf{k})|^2 s_0
  + i\, [\mathbf{d}(\mathbf{k}) \times \mathbf{d}^*(\mathbf{k})] \cdot \mathbf{s} .
\end{equation}

The next order in $\mathcal{H}_{+-}$ consists of a correction~\cite{ABT17, brydon_bogoliubov_2018}
\begin{equation}
\delta H_{N--}(\mathbf{k}) = \frac{1}{2\, |\vec c(\mathbf{k})|}\,
  \Delta_{-+}(\mathbf{k})\, \Delta_{-+}^\dagger(\mathbf{k})
\end{equation}
to the normal-state Hamiltonian $\xi_-(\mathbf{k})\, s_0$. In the pseudospin representation, it can be written as
\begin{equation}
\delta H_{N--}(\mathbf{k}) = \gamma(\mathbf{k})\, s_0 + \mathbf{h}(\mathbf{k}) \cdot \mathbf{s} ,
\end{equation}
where $\gamma(\mathbf{k})$ is a pseudospin-independent correction to the dispersion $\xi_-(\mathbf{k})$ and $\mathbf{h}(\mathbf{k})$ is a pseudomagnetic field coupling to the pseudospin $\mathbf{s}$ \cite{ABT17, brydon_bogoliubov_2018}.  Note that these terms result from \emph{interband} pairing and are proportional to the interband gap product divided by the band splitting $\xi_+(\mathbf{k}) - \xi_-(\mathbf{k}) = 2\, |\vec c(\mathbf{k})|$. A nonzero pseudomagnetic field is only possible if the superconducting pairing breaks TR symmetry. Equation (\ref{eq:normal_dispersion}) shows that this perturbative description is justified as long as the band splitting is large compared to the perturbations due to magnetic field and magnetic order and to the pairing amplitude.

\section{BCS theory with renormalized gap}
\label{app:BCSscaled}

In this appendix, we sketch, using BCS theory for a parabolic band, how a renormalization of the gap at the Fermi energy affects the selfconsistent solution of the gap equation. The internal energy $U$ as a function of the gap amplitude $|\Delta|$ at zero temperature is
\begin{equation}
U \cong N\, \frac{|\Delta|^2}{V} + N \alpha \ln \bigg(\frac{|\Delta|}{\Lambda}\bigg) |\Delta|^2
  + \mathrm{const} ,
\end{equation}
where $N$ is the number of lattice sites, $V > 0$ is the pairing interaction, $\alpha$ is a constant proportional to the density of states \footnote{For a parabolic band with effective mass $m$ and chemical potential $\mu$, one finds $\alpha = v_\mathrm{uc} (2m)^{3/2} \sqrt{\mu}/4\pi^2$, where $v_\mathrm{uc}$ is the unit-cell volume.}, and $\Lambda$ is an ultraviolet cutoff. The first term on the right-hand side stems from mean-field decoupling of the pairing interaction. The second term is the change of energy of the electron system upon opening the gap. Now let us assume that the low-energy gap is rescaled by a factor of $\eta$:
\begin{equation}
U \cong N\, \frac{|\Delta|^2}{V} + N \alpha \ln\!\left(\frac{\eta|\Delta|}{\Lambda}\right) \eta^2|\Delta|^2
  + \mathrm{const} .
\end{equation}
The first term is unchanged since the decoupling is done for the original sublattice-spin Hamiltonian.
The selfconsistent value of $|\Delta|$ is found by minimizing $U$. Solving
\begin{equation}
\frac{dU}{d|\Delta|} \cong 2N\, \frac{|\Delta|}{V} + N \alpha \eta^2 |\Delta|
  + 2N \alpha \eta^2 \ln \bigg(\frac{\eta|\Delta|}{\Lambda}\bigg) |\Delta| = 0
\end{equation}
in terms of $U$ yields
\begin{equation}
|\Delta| \cong \frac{\Lambda}{\eta}\, e^{-1/2} \exp\bigg( {-}\frac{1}{\alpha\eta^2 V} \bigg) .
\end{equation}
In our case, $\eta < 1$. If $\eta$ is close to unity, we obtain a reduction of $|\Delta|$, which is linear in $1-\eta$ (note that $\alpha V$ is small for weak-coupling superconductivity). The more relevant case for our discussion is $\eta$ not close to unity. Then we obtain an \emph{exponential} reduction of $|\Delta|$. Similarly, an exponential reduction of the critical temperature is obtained.

\bibliography{CeRh2As2_GL}

\end{document}